\documentclass[
10pt,
showpacs,preprintnumbers,nofootinbib,
twocolumn,
 amsmath,amssymb,
prl,
groupedaddress,superscriptaddress,
]
{revtex4-1}
\usepackage{graphicx}
\usepackage{bm}
\usepackage[colorlinks=true,urlcolor=blue,citecolor=blue,breaklinks=true]{hyperref}
\usepackage{epsfig}

\newcommand{\cut}[1]{}

\newcommand{\bepg}{$^7$Be($p,\gamma)^8$B }

\begin{document}
\preprint{}

\title{\textit{Ab initio} informed evaluation of the radiative capture of protons on $^7$Be}
\author{K. Kravvaris}
\affiliation{Lawrence Livermore National Laboratory, P.O. Box 808, L-414, Livermore, CA 94551, USA}
\author{P. Navr\'atil}
\affiliation{TRIUMF, 4004 Wesbrook Mall, Vancouver BC, V6T 2A3, Canada}
\author{S. Quaglioni}
\affiliation{Lawrence Livermore National Laboratory, P.O. Box 808, L-414, Livermore, CA 94551, USA}
\author{C.~Hebborn}
\affiliation{Facility for Rare Isotope Beams, East Lansing, MI 48824, USA}
\affiliation{Lawrence Livermore National Laboratory, P.O. Box 808, L-414, Livermore, CA 94551, USA}
\author{G.~Hupin}
\affiliation{Université Paris-Saclay, CNRS/IN2P3, IJCLab, 91405 Orsay, France}
\date{\today}

\begin{abstract}
The radiative capture of protons by $^7$Be, which is the source of $^8$B that $\beta$-decays emitting
the majority of solar neutrinos measured on earth, 
has not yet been measured at astrophysically relevant energies.
The recommended value for its zero-energy S-factor, $S_{17}(0)=20.8\pm(0.7)_\mathrm{exp}\pm(1.4)_\mathrm{theory}$ eV$\cdot$b,  
relies on theoretical extrapolations from higher-energy measurements, 
a process that leads to significant uncertainty. 
We performed a set of first-principle (or, \textit{ab initio}) calculations of the \bepg reaction to provide an independent prediction of the low-energy S-factor with quantified uncertainties.
We demonstrate underlying features in the predicted S-factor allowing the combination of theoretical calculations and measurements to produce an evaluated S-factor of $S_{17}(0)=19.8\pm0.3$ eV$\cdot$b. 
We expect the calculations and uncertainty quantification process described here to set a new 
standard for 
the evaluation of light-ion astrophysical reactions.
\end{abstract}

\maketitle

Astrophysical reactions powering low-mass stars such as our sun have been at the center of 
theoretical and experimental attention ever since nuclear reactions were proposed as a mechanism for nucleosynthesis and energy generation in stellar interiors~\cite{Bethe1939,Burbidge1957}. 
As a result, solar fusion reactions are amongst the most precisely measured and thoroughly evaluated nuclear reactions;
see for example Refs.~\cite{SolarFusion1,SolarFusion2} and references therein.
Occurring at the tail end of the proton-proton chain, the radiative capture of a proton by a $^7$Be nucleus to produce an $^8$B nucleus (or \bepg
reaction) is key in determining the solar neutrino flux measured in terrestrial
observatories~\cite{SuperK,SNOcollaboration}.  Given its importance, it has been measured multiple times over the years
with various techniques~\cite{Filippone1983,Baur1986,Kikuchi1998,Iwasa1999,Davids2001,Baby2003,Schumann2003,Junghans2003,Schumann2006}. 
However, due to Coulomb repulsion between the proton and the $^7$Be nucleus, a direct measurement at the astrophysically
relevant energies is still missing, and theory calculations~\cite{Descouvemont2004,Zhang2015} are used to extrapolate from higher energy experimental data
. 
As a result of this extrapolation process, the uncertainty in the currently recommended~\cite{SolarFusion2} value of the zero-energy S-factor ($S_{17}(0)$),
$S_{17}(0)=20.8\pm0.7\mathrm{(expt)}\pm1.4\mathrm{(theory)}$ eV$\cdot$barn, is dominated by theoretical contributions. 

First-principle (or \textit{ab initio}) theoretical approaches provide an independent prediction of nuclear reaction observables, with the interaction between nucleons being their sole input.
Consequently, the bulk of the theoretical uncertainty of \textit{ab initio} calculations will come from the nuclear interaction employed. 
In this Letter we present first-principle calculations of the $^7$Be+$p$ system, including the \bepg reaction, using nucleon-nucleon (NN) and
three-nucleon (3N) interactions derived from chiral effective field theory ($\chi$EFT)~\cite{Machleidt2011,Epelbaum2015}, with the goal of extracting universal features of the system, and removing (in part) the uncertainty that stems from the choice of a specific interaction and its parametrization. 

The no-core shell model with continuum (NCSMC), first introduced in~\cite{Baroni2013,Baroni2013a}, is a  first-principle technique that has been successful in delivering predictive calculations of nuclear properties of light nuclei by combining bound and dynamic descriptions of an $A$-nucleon system (see Ref.~\cite{Navratil2016} for an in depth review of results). In the NCSMC, the $A$-body Schr{\"o}dinger equation for a total angular momentum $J$ and parity $\pi$ is solved for bound and scattering boundary conditions by means of an ansatz that takes the form
\begin{equation}
|\Psi^{J^\pi}\rangle = \displaystyle\sum_\lambda c_\lambda |A \lambda J^\pi\rangle + \displaystyle\sum_\nu\int r^2 dr \frac{\gamma_\nu^{J^\pi}(r)}{r} \hat{\mathcal{A}}_\nu|\Phi_{\nu r}^{J^\pi}\rangle.
\label{eq:ncsmc}
\end{equation}
The states $|A \lambda J^\pi\rangle$ are obtained from the no-core shell model (NCSM), a many-body harmonic oscillator expansion method~\cite{Barrett2013}, and represent the $\lambda$-th bound-like solution to the $A$-body Schr\"odinger equation. 
The 
channel basis states $\mathcal{A}_\nu|\Phi_{\nu r}^{J^\pi}\rangle$ correspond to totally antisymmetric binary-cluster states where the interacting nuclei (in this case $^7$Be and $p$) are a distance $r$ apart. The collective index $\nu$ corresponds to all asymptotic quantum numbers (internal states, spins, and parities of the fragments, relative angular momentum $\ell$, and spin $s$). 
The unknown discrete parameters $c_\lambda$ and continuous amplitudes $\gamma_\nu^{J^\pi}(r)$ are then determined via the microscopic R-matrix method~\cite{Descouvemont2010}.  
In the NCSMC, the $A$ nucleons are treated as 
point-like, `active', degrees of freedom 
and the same NN+3N interaction 
determines both the intrinsic wave functions (belonging to the projectile, target and aggregate system) obtained in the NCSM, as well as the reaction dynamics. 

Nuclear interactions derived from $\chi$EFT come with a systematic organization in powers of a small expansion parameter $Q<1$. 
The truncation of the expansion at some order gives rise to errors that can be quantified~\cite{Epelbaum2015,Furnstahl2015,Melendez2017,Melendez2019,Wesolowski2021,Ekstrom2019,Kravvaris2020}, by combining consistent calculations performed at different truncation orders. 
In this work we use the NN interactions at next-to-next-to-next-to-leading (4th) order of the chiral expansion defined in Ref.~\cite{Entem2003}, denoted N$^3$LO$^*$, and those at 3rd, 4th and 5th order of Ref. \cite{Entem2017}, denoted respectively N$^2$LO, N$^3$LO, and N$^4$LO. 
These NN interactions are supplemented by the 3N interaction of Ref.~\cite{VanKolck94} with both local (3N$_\mathrm{loc}$)~\cite{Navratil2007,Gazit2019} and local plus non-local (3N$_\mathrm{lnl}$)  regulators~\cite{Soma2020,Kravvaris2020}. 
Finally, a 3N interaction employing  a local plus non-local regulator but with an added sub-leading contact term enhancing the strength of the spin-orbit interaction ($E_7$ term), as described in \cite{Girlanda2011}, was also employed (3N$_\mathrm{lnl}^*$). 
A list of all NN+3N combinations that are employed in this work can be found in Table~\ref{tab:ANCSfactors}. 
The interactions described in Ref.~\cite{Entem2017} 
are a consistent set and we expect truncation errors extracted from their differences to be similar to errors for other parametrizations.  

\begin{figure}
\centering
\includegraphics[width=0.47\textwidth]{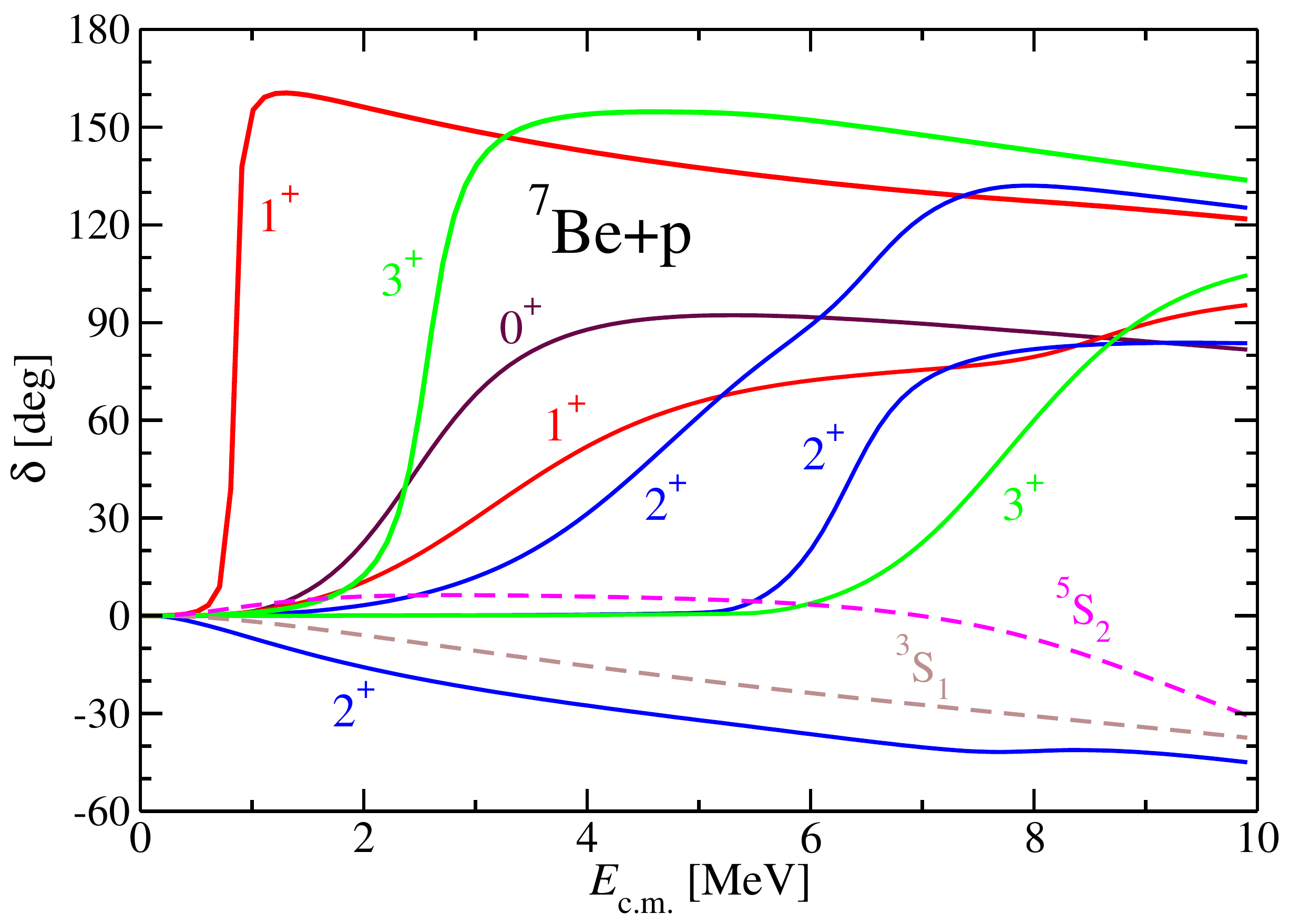}
\caption{$^7$Be+$p$ eigenphase shifts (solid lines) and $^3S_1$ and $^5S_2$ diagonal phase shifts (dashed lines) obtained from the NCSMC approach with the N$^4$LO+3N$^*_{\rm lnl}$ interaction. } 
\label{fig:phaseshift}
\end{figure}


In all calculations, the interaction was softened using a similarity renormalization group (SRG) transformation ~\cite{Jurgenson2009}, 
including induced forces up to the three-body level. Four- and higher-body induced terms are small at the $\lambda_{\mathrm{SRG}}{=}2.0$ fm$^{-1}$ resolution scale used in present calculations~\cite{PhysRevC.103.035801}. 
The NCSM calculations of the $^7$Be and $^8$B nuclei were carried out allowing for up to nine quanta of excitation ($N_\mathrm{max}=9$) for natural parity states and $N_\mathrm{max}=10$ for the negative parity states of $^8$B. The relative motion between the two fragments was also computed within the same $N_\mathrm{max}=9(10)$ space for positive (negative) parity channels.
The harmonic oscillator parameter $\hbar\Omega$ was chosen at 20 MeV, which is close to the value that minimizes the ground-state energies of investigated nuclei~\cite{Navratil2011a}. 
The reaction channel basis states are constructed by taking into account the first five states of $^7$Be ($J^\pi=3/2^-,1/2^-,7/2^-,5/2^-,5/2^-$). Earlier work~\cite{Navratil2011} has demonstrated that this choice is sufficient to reach convergence in the channel basis expansion.  The discrete part of the NCSMC ansatz is spanned by the ten lowest energy states of $^8$B for each parity value, with spin values ranging from $J=0$ to $J=4$. 

The positive parity eigenphase shifts for $^7$Be+$p$ scattering obtained using the N$^4$LO+3N$^*_{\rm lnl}$ chiral interaction up to 10 MeV in center of mass energy ($E_\mathrm{c.m.}$) show the well-established $1^+$ and $3^+$ $^8$B resonances as well as 
several other, yet unobserved, broader resonances (Fig.~\ref{fig:phaseshift}). 
The NCSMC $S$-wave phase shifts manifest scattering length signs consistent with those determined in recent measurements (negative for $^5S_2$, positive for $^3S_1$)~\cite{PhysRevC.99.045807}, unlike previous calculations~\cite{Navratil2011a} that disregarded the discrete portion of the ansatz of Eq.~(\ref{eq:ncsmc}). 
Using this parametrization of the NN+3N interaction, the 
$2^+$ state in $^8$B 
is slightly unbound. The very narrow near-threshold $2^+$ resonance is not visible in the figure. 
We find it is difficult to produce a bound $^8$B ground state, with the exception of the N$^3$LO+3N$_\mathrm{loc}$ interaction, which reproduces the ground state energy at the 30 keV level.
Owing to the parity difference between the ground state of $^8$B ($2^+$) and that of $^7$Be ($3/2^-$), the low-energy cross section of the \bepg radiative capture proceeds via an E1 transition~\cite{Robertson1973}.
While the bare E1 operator has a one-body form, a consistent SRG evolution to the same scale as the interaction, will induce many-body contributions. 
These contributions were found to be of a short-range nature~\cite{Schuster2014}, thus we expect them to be small in a calculation involving a loosely-bound system such as $^8$B. 
Operators of the M1/E2 types contribute at higher energies and are treated using a closure approximation of the $^8$B NCSM states and $^7$Be+$p$ channels ~\cite{Navratil2016}. 


The \bepg S-factor is sensitive to the spatial extent of the $^8$B wave function~\cite{PhysRevC.61.025801}, that is in turn determined by the ground state binding energy of 136(1) keV~\cite{Audi2012}. 
While it is impossible to exactly reproduce this binding energy in all NN+3N interaction models used in this study, especially when considering all sources of uncertainty, it is possible to adjust the NCSMC ansatz so that the overall binding energy reproduces the experimental value, as done for example in~\cite{Hupin2019}. 
This phenomenological correction, dubbed NCSMC-pheno, is performed by shifting the NCSM eigenenergies of $^7$Be so that the excitation energies (and therefore thresholds) match the experimental ones exactly. 
This 
ensures that decaying states have the correct phase space available, corresponding to their energy.
Furthermore, the $^8$B NCSM eigenenergies in the $2^+,1^+,3^+$ channels are also modified bringing the bound and unbound NCSMC states in the experimentally observed positions. The resulting 
scattering lengths, asymptotic normalization coefficients (ANCs), and $S_{17}(0)$ are shown in Table~\ref{tab:ANCSfactors}.
We consider this approach to be an \textit{ab initio} guided evaluation process where experimental data are fed into the theoretical prediction to correct small deficiencies of the nuclear interaction, resulting in greater predictive capability. 

\begin{table}[t]
\centering
\begin{tabular}{c@{\hskip 0.1in} | r@{\hskip 0.05in}c@{\hskip 0.05in}c@{\hskip 0.05in}c@{\hskip 0.0in}c}
  & $C_{p_{1/2}}$ & $C_{p_{3/2}}$ & $a_{1}$ & $a_{2}$& $S_{17}(0)$\\ 
 \hline
 \hline
 N$^2$LO+3N$_\mathrm{lnl}$      & 0.384 & 0.691 & 4.4(1) & -0.5(1) & 23.9 \\  
 N$^3$LO+3N$_\mathrm{lnl}$      & 0.390 & 0.678 & 1.3(1) & -4.7(1) & 23.5 \\  
 N$^4$LO+3N$_\mathrm{lnl}$      & 0.354 & 0.669 & 1.6(1) & -4.4(1) & 22.0 \\  
 N$^4$LO+3N$^*_\mathrm{lnl}$    & 0.343 & 0.621 & 1.3(1) & -5.0(1) & 19.3 \\  
 N$^3$LO$^*$+3N$_\mathrm{lnl}$  & 0.334 & 0.663 & 0.1(1) & -7.7(1) & 21.1 \\  
 N$^3$LO$^*$+3N$_\mathrm{loc}$  & 0.308 & 0.584 & 2.5(1) & -3.6(2) & 16.8 \\ 
 Ref.~\cite{PhysRevC.99.045807} & 0.315(9) & 0.66(2) & 17.34$^{+1.11}_{-1.33}$ & -3.18$^{+0.55}_{-0.50}$ &  \\
 \hline
 \end{tabular}
\caption{Values for asymptotic normalization coefficients (ANCs) in fm$^{-1/2}$, scattering lengths in fm, and zero-energy S-factor in eV$\cdot$b obtained from the set of interactions used in this work, after applying a phenomenological correction (see text) to the $^8$B bound-state as well as the $1^+$ and $3^+$ resonance energies. The experimentally extracted values from Ref.~\cite{PhysRevC.99.045807} are also included for comparison.}
 \label{tab:ANCSfactors}
 \end{table}

\begin{figure}[t]
\centering
\includegraphics[width=0.47\textwidth]{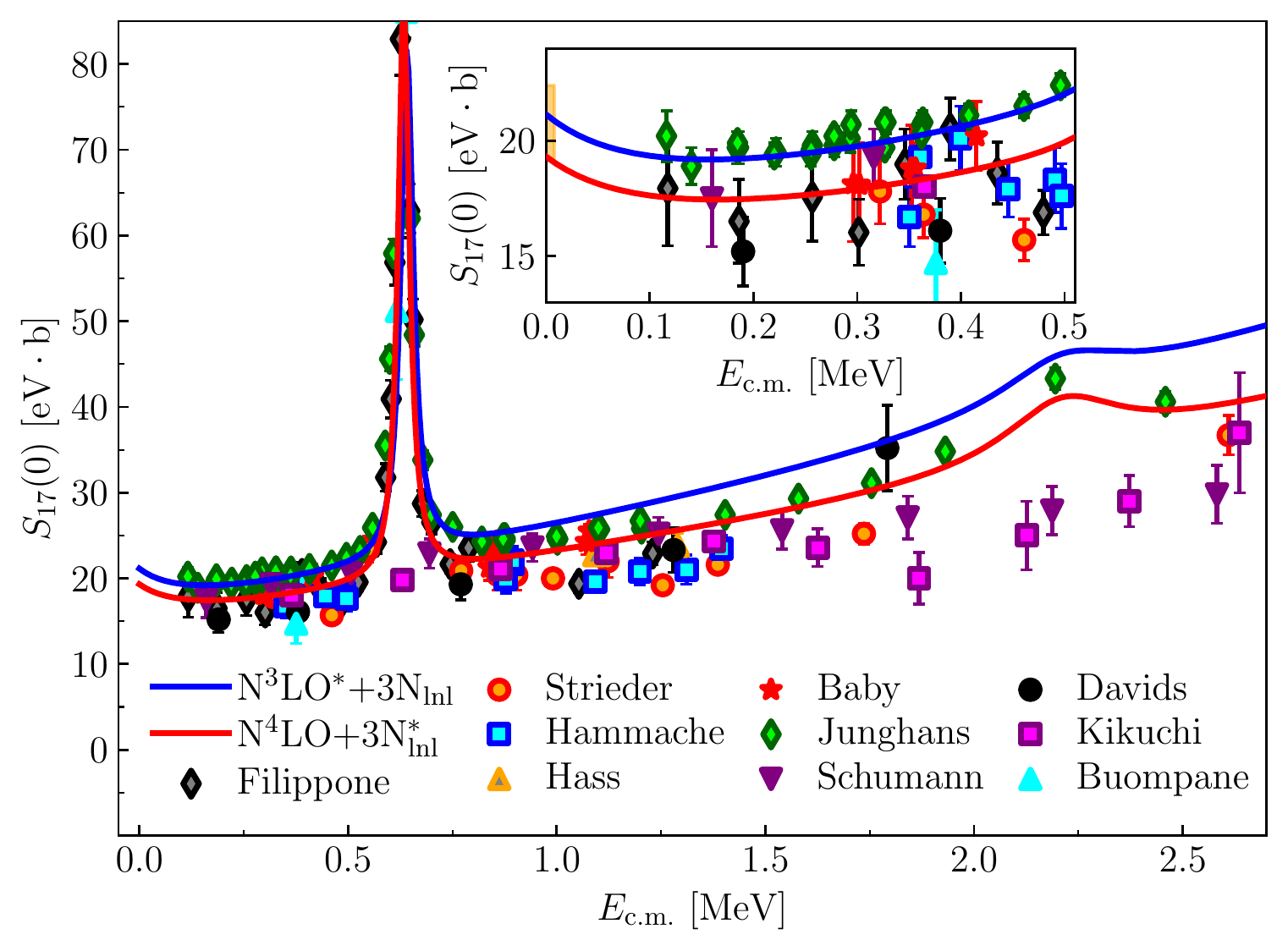}
\caption{Astrophysical S-factor of the \bepg radiative capture obtained from the NCSMC-pheno approach with the N$^3$LO$^*$+3N$_{\rm lnl}$ (blue line) and the N$^4$LO+3N$^*_{\rm lnl}$ (red line) interactions compared to experimental data. The inset shows the low-energy part and shows the evaluation of $S_{17}(0)$ with its uncertainty~\cite{SolarFusion2} (orange box).} 
\label{fig:sfactor}
\end{figure}

The astrophysical S-factors obtained within this NCSMC-pheno approach starting from the N$^3$LO$^*$+3N$_{\rm lnl}$ and the N$^4$LO+3N$^*_{\rm lnl}$ interactions reproduce the experimentally observed sharp peak at about $0.6$~MeV and, to a lesser extent, the broader structure around $2.2$~MeV that are caused by magnetic dipole (M1) and electric quadrupole (E2) transitions of, respectively, the $1^+$ and $3^+$ resonances to the $2^+$ ground state of $^8$B (Fig.~\ref{fig:sfactor}).
On the one hand, the calculation using the N$^4$LO+3N$^*_{\rm lnl}$ interaction matches well the direct measurement from Junghans et al.~\cite{PhysRevC.68.065803} starting at the $1^+$ resonance and up until the $3^+$ bump, but slightly underestimates them at low energies. 
On the other hand, the N$^3$LO$^*$+3N$_{\rm lnl}$ results follow the Junghans et al.\ data from threshold and up to the $1^+$ peak, 
but overestimate them somewhat at higher energies. 

\begin{figure*}[t]
\centering
  \includegraphics[width=0.8\textwidth]{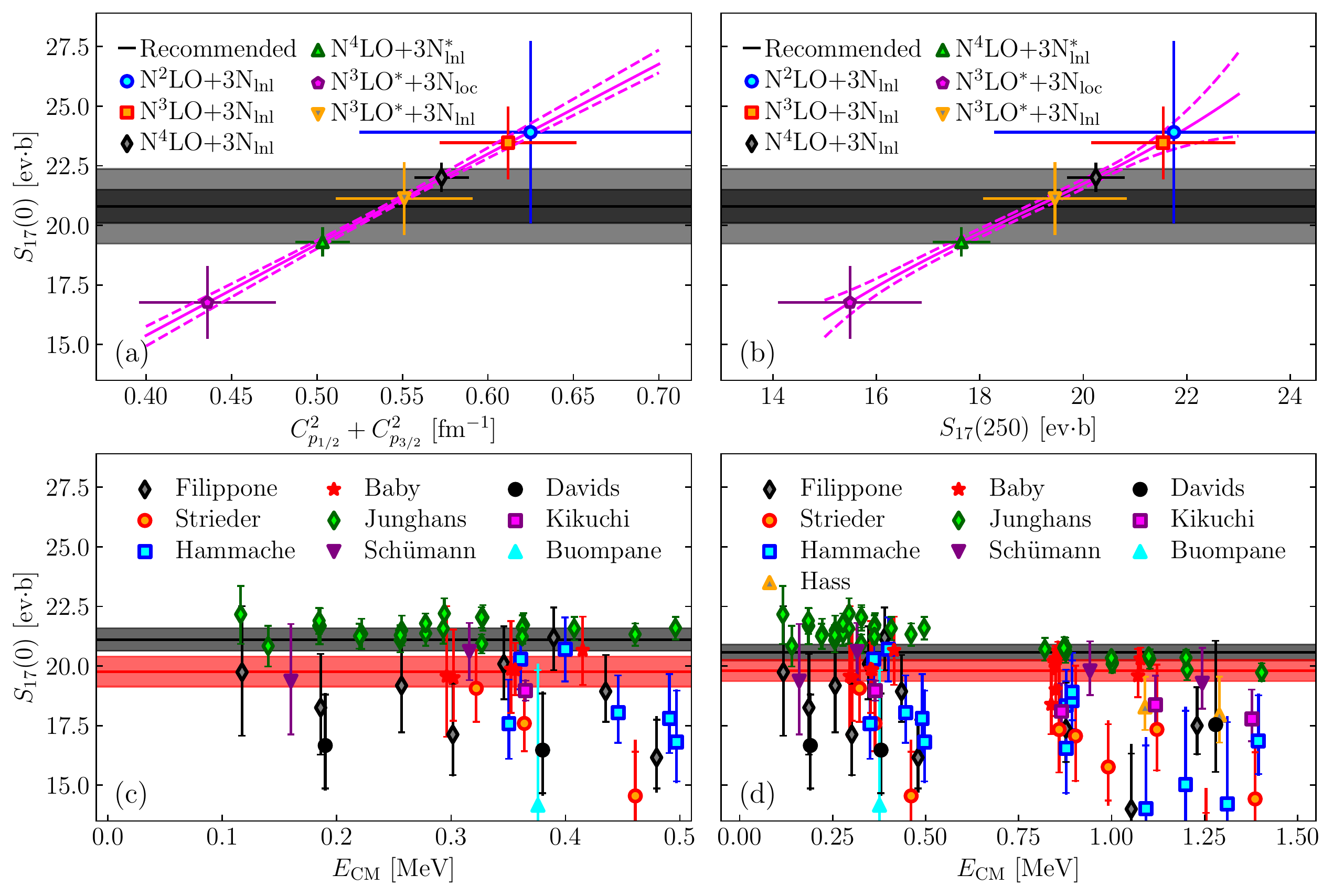}
  \caption{(a) S-factor at zero energy versus the sum of the squares of the ANCs for all interactions used in this work. Theoretical uncertainties are estimated using $Q=0.4$. The linear fit (magenta line), along with its $1\sigma$ uncertainty band (dashed lines), suggests a reduced theoretical uncertainty compared to the current evaluation (grey shaded area, dark grey band for experimental uncertainty only). (b) Correlation between the S-factor at 0 keV and at 250 keV. The fitted line is a cubic polynomial (see text for details). The $1\sigma$ uncertainty of the fit is significantly smaller at the region of interest compared to the edges. (c) Resulting values for $S_{17}(0)$ using the regression method described in the text for each data point with $E_\mathrm{c.m.}\leq0.5$ MeV. Error bars correspond to $1\sigma$ (theoretical plus experimental) uncertainty, while the cap points denote the experimental uncertainty only. The black line corresponds to a best fit value for $S_{17}(0)$ with a 3$\sigma$ uncertainty band. The red line uses a modified fit measure (see text for details). (d) Same as (c) but using also data with $0.8\leq E_\mathrm{c.m.}\leq 1.5$ MeV.}
    \label{fig:sffit}
\end{figure*}

More in general, none of the calculations we carried out is able to describe the S-factor experimental data over the entire range of energies.  However, 
The multiple calculations of the $^7$Be + $p$ system allow for a more systematic look at its inherent properties without focusing on a specific interaction. 
Indeed, the use of various chiral order truncations and differing regulators gives us a window into the universal properties of the system as described by $\chi$EFT.
As 
previously pointed out in Refs.~\cite{Xu1994,Baye2000b}, 
there is a linear relationship 
between $S_{17}(0)$ and the sum of the squares of the ANCs ($C_{p_{1/2}}^2 + C_{p_{3/2}}^2$). 
We observe such a relationship (Fig.~\ref{fig:sffit}a), with all NCSMC calculations lying along a line with slope 38.53 $\pm$ 1.45 eV$\cdot$b$\cdot$fm. 
The error bars in the points of Figs.~\ref{fig:sffit}a,b, propagated through the linear regression, correspond to an estimate of the chiral truncation uncertainty for each interaction. 
We note that the error bars shown for the theoretical calculations in Fig.~\ref{fig:sffit}a should not be treated as uncorrelated errors; the relation between the ANCs and $S_{17}(0)$ is inherent in the equations being solved, not the specific interaction. 
The tight uncertainty band on the fitted line implies that if ANCs were accurately measured, the resulting theoretical uncertainty would be orders of magnitude smaller than the currently recommended value.
Nevertheless, we can still look for correlations with already available experimental data, that could lead to an accurate determination of $S_{17}(0)$. 

We also find a 
nearly linear correlation between $S_{17}(0)$ and the S-factor at energies $E_\mathrm{c.m.}<0.5$ MeV. 
As an example, in Fig~\ref{fig:sffit}b we show the correlation between the S-factor at zero and at 250 keV;
the fit is obtained with a cubic polynomial incorporating deviations that are bound to exist when comparing different energies. We verified that the degree of the polynomial used does not significantly impact our results. 
While the $1\sigma$ uncertainty of the fit is slightly larger than in the linear fit of the ANCs,  it is still an order of magnitude smaller than the recommended theoretical uncertainty in the region of the 
evaluation of Ref.~\cite{SolarFusion2}. 

We 
combine 
the tight relationship we have identified between the S-factor at zero and higher energies with experimental measurements for energies $E_\mathrm{c.m.}\leq0.5$ MeV to provide an $\textit{evaluated}$ prediction for $S_{17}(0)$ 
that includes both theoretical and experimental uncertainties. 
Specifically, we $i)$ carry out a regression of the predicted $S_{17}(E_{\rm c.m.})$ versus $S_{17}(0)$ relationship for each $E_{\rm c.m.}$ value where an experimental measurement exists 
(similar to Fig.~\ref{fig:sffit}b); and $ii)$  
infer the $S_{17}(0)$ (ordinate) corresponding to the measured $S_{17}(E_{\rm c.m.})$ (abscissa) from the fitted polynomials (Fig.~\ref{fig:sffit}c). The calculations at N$^4$LO are essential in reducing 
the regression uncertainty, resulting in a small theoretical error bar for the 
inferred $S_{17}(0)$. 
We find the experimental contribution dominates the uncertainty budget, with the theoretical part being about an order of magnitude smaller. 
A regression to a horizontal line using all inferred data points yields $S_{17}(0) = 21.1 \pm 0.2$ eV$\cdot$b (Fig.~\ref{fig:sffit}c), which agrees with the currently recommended value within 1$\sigma$. Extending this protocol to include energies with 0.8 MeV $\leq E_\mathrm{c.m.} \leq 1.5$ MeV brings the evaluation down to $20.4\pm0.1$ eV$\cdot$b (Fig.~\ref{fig:sffit}d). However, we find a discrepancy between the $S_{17}(0)$ values predicted with the lower- and higher-energy data from Junghans et al.~\cite{Junghans2003} (green diamonds). This discrepancy can be further illustrated by using only the higher-energy data for the fit, in which case we obtain $S_{17}(0)$=20.2$\pm$0.1 eV$\cdot$b.

The discrepancy in the values of $S_{17}(0)$ extracted from the three regressions described above can be traced back to our treatment of all measurements in an experiment as uncorrelated. This treatment leads to fits dominated by a single experiment with multiple measurements and small error bars, as the one of Ref.~\cite{Junghans2003}. This no correlation assumption is not necessarily true when looking at, for example, data points at nearby energies. 
Indeed, computing the covariance matrix for the theoretical calculations at various energies, we find large correlation coefficients ($>$0.9) across the full energy range in question. 
While, there are a few 
options (from guessing a covariance form, to using theoretical covariances), directly addressing this issue   
goes beyond the scope of the present work. Instead, we assign a factor normalizing the contribution of each experiment based on the number of data points. 
Such a normalization is equivalent to a assuming a fully correlated covariance matrix, where only one `average' data point is used from each experiment. 
Following this process to fit to the $E_\mathrm{c.m.}\leq0.5$ MeV data yields $S_{17}(0) = 19.8 \pm 0.2$ eV$\cdot$b, which again agrees with the recommended value within uncertainties, but disagrees with the value obtained from the fit that did not consider covariances at the 3$\sigma$ level (Fig.~\ref{fig:sffit}c). 
The fit to the $0.8\leq E_\mathrm{\rm c.m.}\leq 1.5$ MeV data gives a value of 19.8$\pm$0.2 eV$\cdot$b, while fitting over both energy ranges yields 19.8$\pm$0.1 eV$\cdot$b (Fig.~\ref{fig:sffit}d). 
Based on the robustness of the fit with modified experimental weights
we arrive at a suggested  value for the $^7$Be$(p,\gamma)^8$B S-factor at zero energy of 19.8$\pm$0.3 eV$\cdot$b, with the 1$\sigma$ uncertainty 
conservatively increased to account for the different 
data fitting procedures. 


In conclusion, we have conducted an extensive \textit{ab initio} analysis of the $^7$Be+$p$ system, providing theoretical predictions for phase shifts and the radiative capture cross section. 
To this end, we performed NCSMC calculations with a variety of $\chi$EFT interactions, all of which provide a reasonable description of the $^7$Be and $^8$B nuclei, and further refined them to reproduce observed binding energies and resonance positions.  Finally, we introduced a new protocol that combines predictive \textit{ab initio} calculations (with quantified uncertainties) and experimental data to 
evaluate observables in regions where experimental measurements are not feasible, and demonstrated the necessity for high-order (larger or equal to fifth order) calculations with reduced chiral truncation uncertainties. 
Alongside such calculations, high-precision experiments both at low and higher energies can help in further pinning down the value of $S_{17}(0)$.
We believe that this work sets a new standard for 
the evaluation of thermonuclear reaction rates that will lead to more robust predictive capabilities for astrophysical models. It also highlights the necessity of incorporating estimates of experimental and theoretical covariances in the evaluation process.


\paragraph{}\begin{acknowledgments}
Computing support for this work came from the LLNL institutional Computing Grand Challenge
program and from an INCITE Award on the Summit supercomputer of the Oak Ridge Leadership Computing Facility
(OLCF) at ORNL.
Prepared in part by LLNL under Contract DE-AC52-07NA27344. 
This material is based upon work supported by the U.S.\ Department of Energy, Office of Science, Office of Nuclear Physics, 
under Work Proposal No.\ SCW0498, LLNL LDRD Project No. 20-LW-046, under the FRIB Theory Alliance award no. DE-SC0013617, and by the NSERC Grants SAPIN-2016-00033 and SAPPJ-2019-00039.  
TRIUMF receives federal funding via a contribution agreement with the National Research Council of Canada. 
\end{acknowledgments}


\begin{thebibliography}{50}%
\makeatletter
\providecommand \@ifxundefined [1]{%
 \@ifx{#1\undefined}
}%
\providecommand \@ifnum [1]{%
 \ifnum #1\expandafter \@firstoftwo
 \else \expandafter \@secondoftwo
 \fi
}%
\providecommand \@ifx [1]{%
 \ifx #1\expandafter \@firstoftwo
 \else \expandafter \@secondoftwo
 \fi
}%
\providecommand \natexlab [1]{#1}%
\providecommand \enquote  [1]{``#1''}%
\providecommand \bibnamefont  [1]{#1}%
\providecommand \bibfnamefont [1]{#1}%
\providecommand \citenamefont [1]{#1}%
\providecommand \href@noop [0]{\@secondoftwo}%
\providecommand \href [0]{\begingroup \@sanitize@url \@href}%
\providecommand \@href[1]{\@@startlink{#1}\@@href}%
\providecommand \@@href[1]{\endgroup#1\@@endlink}%
\providecommand \@sanitize@url [0]{\catcode `\\12\catcode `\$12\catcode
  `\&12\catcode `\#12\catcode `\^12\catcode `\_12\catcode `\%12\relax}%
\providecommand \@@startlink[1]{}%
\providecommand \@@endlink[0]{}%
\providecommand \url  [0]{\begingroup\@sanitize@url \@url }%
\providecommand \@url [1]{\endgroup\@href {#1}{\urlprefix }}%
\providecommand \urlprefix  [0]{URL }%
\providecommand \Eprint [0]{\href }%
\providecommand \doibase [0]{https://doi.org/}%
\providecommand \selectlanguage [0]{\@gobble}%
\providecommand \bibinfo  [0]{\@secondoftwo}%
\providecommand \bibfield  [0]{\@secondoftwo}%
\providecommand \translation [1]{[#1]}%
\providecommand \BibitemOpen [0]{}%
\providecommand \bibitemStop [0]{}%
\providecommand \bibitemNoStop [0]{.\EOS\space}%
\providecommand \EOS [0]{\spacefactor3000\relax}%
\providecommand \BibitemShut  [1]{\csname bibitem#1\endcsname}%
\let\auto@bib@innerbib\@empty
\bibitem [{\citenamefont {Bethe}(1939)}]{Bethe1939}%
  \BibitemOpen
  \bibfield  {author} {\bibinfo {author} {\bibfnamefont {H.~A.}\ \bibnamefont
  {Bethe}},\ }\bibfield  {title} {\bibinfo {title} {Energy production in
  stars},\ }\href {https://doi.org/10.1103/PhysRev.55.434} {\bibfield
  {journal} {\bibinfo  {journal} {Phys. Rev.}\ }\textbf {\bibinfo {volume}
  {55}},\ \bibinfo {pages} {434} (\bibinfo {year} {1939})}\BibitemShut
  {NoStop}%
\bibitem [{\citenamefont {Burbidge}\ \emph {et~al.}(1957)\citenamefont
  {Burbidge}, \citenamefont {Burbidge}, \citenamefont {Fowler},\ and\
  \citenamefont {Hoyle}}]{Burbidge1957}%
  \BibitemOpen
  \bibfield  {author} {\bibinfo {author} {\bibfnamefont {E.~M.}\ \bibnamefont
  {Burbidge}}, \bibinfo {author} {\bibfnamefont {G.~R.}\ \bibnamefont
  {Burbidge}}, \bibinfo {author} {\bibfnamefont {W.~A.}\ \bibnamefont
  {Fowler}},\ and\ \bibinfo {author} {\bibfnamefont {F.}~\bibnamefont
  {Hoyle}},\ }\bibfield  {title} {\bibinfo {title} {Synthesis of the elements
  in stars},\ }\href {https://doi.org/10.1103/RevModPhys.29.547} {\bibfield
  {journal} {\bibinfo  {journal} {Rev. Mod. Phys.}\ }\textbf {\bibinfo {volume}
  {29}},\ \bibinfo {pages} {547} (\bibinfo {year} {1957})}\BibitemShut
  {NoStop}%
\bibitem [{\citenamefont {Adelberger}\ \emph {et~al.}(1998)\citenamefont
  {Adelberger}, \citenamefont {Austin}, \citenamefont {Bahcall}, \citenamefont
  {Balantekin}, \citenamefont {Bogaert}, \citenamefont {Brown}, \citenamefont
  {Buchmann}, \citenamefont {Cecil}, \citenamefont {Champagne}, \citenamefont
  {de~Braeckeleer}, \citenamefont {Duba}, \citenamefont {Elliott},
  \citenamefont {Freedman}, \citenamefont {Gai}, \citenamefont {Goldring},
  \citenamefont {Gould}, \citenamefont {Gruzinov}, \citenamefont {Haxton},
  \citenamefont {Heeger}, \citenamefont {Henley}, \citenamefont {Johnson},
  \citenamefont {Kamionkowski}, \citenamefont {Kavanagh}, \citenamefont
  {Koonin}, \citenamefont {Kubodera}, \citenamefont {Langanke}, \citenamefont
  {Motobayashi}, \citenamefont {Pandharipande}, \citenamefont {Parker},
  \citenamefont {Robertson}, \citenamefont {Rolfs}, \citenamefont {Sawyer},
  \citenamefont {Shaviv}, \citenamefont {Shoppa}, \citenamefont {Snover},
  \citenamefont {Swanson}, \citenamefont {Tribble}, \citenamefont
  {Turck-Chi\`eze},\ and\ \citenamefont {Wilkerson}}]{SolarFusion1}%
  \BibitemOpen
  \bibfield  {author} {\bibinfo {author} {\bibfnamefont {E.~G.}\ \bibnamefont
  {Adelberger}}, \bibinfo {author} {\bibfnamefont {S.~M.}\ \bibnamefont
  {Austin}}, \bibinfo {author} {\bibfnamefont {J.~N.}\ \bibnamefont {Bahcall}},
  \bibinfo {author} {\bibfnamefont {A.~B.}\ \bibnamefont {Balantekin}},
  \bibinfo {author} {\bibfnamefont {G.}~\bibnamefont {Bogaert}}, \bibinfo
  {author} {\bibfnamefont {L.~S.}\ \bibnamefont {Brown}}, \bibinfo {author}
  {\bibfnamefont {L.}~\bibnamefont {Buchmann}}, \bibinfo {author}
  {\bibfnamefont {F.~E.}\ \bibnamefont {Cecil}}, \bibinfo {author}
  {\bibfnamefont {A.~E.}\ \bibnamefont {Champagne}}, \bibinfo {author}
  {\bibfnamefont {L.}~\bibnamefont {de~Braeckeleer}}, \bibinfo {author}
  {\bibfnamefont {C.~A.}\ \bibnamefont {Duba}}, \bibinfo {author}
  {\bibfnamefont {S.~R.}\ \bibnamefont {Elliott}}, \bibinfo {author}
  {\bibfnamefont {S.~J.}\ \bibnamefont {Freedman}}, \bibinfo {author}
  {\bibfnamefont {M.}~\bibnamefont {Gai}}, \bibinfo {author} {\bibfnamefont
  {G.}~\bibnamefont {Goldring}}, \bibinfo {author} {\bibfnamefont {C.~R.}\
  \bibnamefont {Gould}}, \bibinfo {author} {\bibfnamefont {A.}~\bibnamefont
  {Gruzinov}}, \bibinfo {author} {\bibfnamefont {W.~C.}\ \bibnamefont
  {Haxton}}, \bibinfo {author} {\bibfnamefont {K.~M.}\ \bibnamefont {Heeger}},
  \bibinfo {author} {\bibfnamefont {E.}~\bibnamefont {Henley}}, \bibinfo
  {author} {\bibfnamefont {C.~W.}\ \bibnamefont {Johnson}}, \bibinfo {author}
  {\bibfnamefont {M.}~\bibnamefont {Kamionkowski}}, \bibinfo {author}
  {\bibfnamefont {R.~W.}\ \bibnamefont {Kavanagh}}, \bibinfo {author}
  {\bibfnamefont {S.~E.}\ \bibnamefont {Koonin}}, \bibinfo {author}
  {\bibfnamefont {K.}~\bibnamefont {Kubodera}}, \bibinfo {author}
  {\bibfnamefont {K.}~\bibnamefont {Langanke}}, \bibinfo {author}
  {\bibfnamefont {T.}~\bibnamefont {Motobayashi}}, \bibinfo {author}
  {\bibfnamefont {V.}~\bibnamefont {Pandharipande}}, \bibinfo {author}
  {\bibfnamefont {P.}~\bibnamefont {Parker}}, \bibinfo {author} {\bibfnamefont
  {R.~G.~H.}\ \bibnamefont {Robertson}}, \bibinfo {author} {\bibfnamefont
  {C.}~\bibnamefont {Rolfs}}, \bibinfo {author} {\bibfnamefont {R.~F.}\
  \bibnamefont {Sawyer}}, \bibinfo {author} {\bibfnamefont {N.}~\bibnamefont
  {Shaviv}}, \bibinfo {author} {\bibfnamefont {T.~D.}\ \bibnamefont {Shoppa}},
  \bibinfo {author} {\bibfnamefont {K.~A.}\ \bibnamefont {Snover}}, \bibinfo
  {author} {\bibfnamefont {E.}~\bibnamefont {Swanson}}, \bibinfo {author}
  {\bibfnamefont {R.~E.}\ \bibnamefont {Tribble}}, \bibinfo {author}
  {\bibfnamefont {S.}~\bibnamefont {Turck-Chi\`eze}},\ and\ \bibinfo {author}
  {\bibfnamefont {J.~F.}\ \bibnamefont {Wilkerson}},\ }\bibfield  {title}
  {\bibinfo {title} {Solar fusion cross sections},\ }\href
  {https://doi.org/10.1103/RevModPhys.70.1265} {\bibfield  {journal} {\bibinfo
  {journal} {Rev. Mod. Phys.}\ }\textbf {\bibinfo {volume} {70}},\ \bibinfo
  {pages} {1265} (\bibinfo {year} {1998})}\BibitemShut {NoStop}%
\bibitem [{\citenamefont {Adelberger}\ \emph {et~al.}(2011)\citenamefont
  {Adelberger}, \citenamefont {Garc\'{\i}a}, \citenamefont {Robertson},
  \citenamefont {Snover}, \citenamefont {Balantekin}, \citenamefont {Heeger},
  \citenamefont {Ramsey-Musolf}, \citenamefont {Bemmerer}, \citenamefont
  {Junghans}, \citenamefont {Bertulani}, \citenamefont {Chen}, \citenamefont
  {Costantini}, \citenamefont {Prati}, \citenamefont {Couder}, \citenamefont
  {Uberseder}, \citenamefont {Wiescher}, \citenamefont {Cyburt}, \citenamefont
  {Davids}, \citenamefont {Freedman}, \citenamefont {Gai}, \citenamefont
  {Gazit}, \citenamefont {Gialanella}, \citenamefont {Imbriani}, \citenamefont
  {Greife}, \citenamefont {Hass}, \citenamefont {Haxton}, \citenamefont
  {Itahashi}, \citenamefont {Kubodera}, \citenamefont {Langanke}, \citenamefont
  {Leitner}, \citenamefont {Leitner}, \citenamefont {Vetter}, \citenamefont
  {Winslow}, \citenamefont {Marcucci}, \citenamefont {Motobayashi},
  \citenamefont {Mukhamedzhanov}, \citenamefont {Tribble}, \citenamefont
  {Nollett}, \citenamefont {Nunes}, \citenamefont {Park}, \citenamefont
  {Parker}, \citenamefont {Schiavilla}, \citenamefont {Simpson}, \citenamefont
  {Spitaleri}, \citenamefont {Strieder}, \citenamefont {Trautvetter},
  \citenamefont {Suemmerer},\ and\ \citenamefont {Typel}}]{SolarFusion2}%
  \BibitemOpen
  \bibfield  {author} {\bibinfo {author} {\bibfnamefont {E.~G.}\ \bibnamefont
  {Adelberger}}, \bibinfo {author} {\bibfnamefont {A.}~\bibnamefont
  {Garc\'{\i}a}}, \bibinfo {author} {\bibfnamefont {R.~G.~Hamish}\ \bibnamefont
  {Robertson}}, \bibinfo {author} {\bibfnamefont {K.~A.}\ \bibnamefont
  {Snover}}, \bibinfo {author} {\bibfnamefont {A.~B.}\ \bibnamefont
  {Balantekin}}, \bibinfo {author} {\bibfnamefont {K.}~\bibnamefont {Heeger}},
  \bibinfo {author} {\bibfnamefont {M.~J.}\ \bibnamefont {Ramsey-Musolf}},
  \bibinfo {author} {\bibfnamefont {D.}~\bibnamefont {Bemmerer}}, \bibinfo
  {author} {\bibfnamefont {A.}~\bibnamefont {Junghans}}, \bibinfo {author}
  {\bibfnamefont {C.~A.}\ \bibnamefont {Bertulani}}, \bibinfo {author}
  {\bibfnamefont {J.-W.}\ \bibnamefont {Chen}}, \bibinfo {author}
  {\bibfnamefont {H.}~\bibnamefont {Costantini}}, \bibinfo {author}
  {\bibfnamefont {P.}~\bibnamefont {Prati}}, \bibinfo {author} {\bibfnamefont
  {M.}~\bibnamefont {Couder}}, \bibinfo {author} {\bibfnamefont
  {E.}~\bibnamefont {Uberseder}}, \bibinfo {author} {\bibfnamefont
  {M.}~\bibnamefont {Wiescher}}, \bibinfo {author} {\bibfnamefont
  {R.}~\bibnamefont {Cyburt}}, \bibinfo {author} {\bibfnamefont
  {B.}~\bibnamefont {Davids}}, \bibinfo {author} {\bibfnamefont {S.~J.}\
  \bibnamefont {Freedman}}, \bibinfo {author} {\bibfnamefont {M.}~\bibnamefont
  {Gai}}, \bibinfo {author} {\bibfnamefont {D.}~\bibnamefont {Gazit}}, \bibinfo
  {author} {\bibfnamefont {L.}~\bibnamefont {Gialanella}}, \bibinfo {author}
  {\bibfnamefont {G.}~\bibnamefont {Imbriani}}, \bibinfo {author}
  {\bibfnamefont {U.}~\bibnamefont {Greife}}, \bibinfo {author} {\bibfnamefont
  {M.}~\bibnamefont {Hass}}, \bibinfo {author} {\bibfnamefont {W.~C.}\
  \bibnamefont {Haxton}}, \bibinfo {author} {\bibfnamefont {T.}~\bibnamefont
  {Itahashi}}, \bibinfo {author} {\bibfnamefont {K.}~\bibnamefont {Kubodera}},
  \bibinfo {author} {\bibfnamefont {K.}~\bibnamefont {Langanke}}, \bibinfo
  {author} {\bibfnamefont {D.}~\bibnamefont {Leitner}}, \bibinfo {author}
  {\bibfnamefont {M.}~\bibnamefont {Leitner}}, \bibinfo {author} {\bibfnamefont
  {P.}~\bibnamefont {Vetter}}, \bibinfo {author} {\bibfnamefont
  {L.}~\bibnamefont {Winslow}}, \bibinfo {author} {\bibfnamefont {L.~E.}\
  \bibnamefont {Marcucci}}, \bibinfo {author} {\bibfnamefont {T.}~\bibnamefont
  {Motobayashi}}, \bibinfo {author} {\bibfnamefont {A.}~\bibnamefont
  {Mukhamedzhanov}}, \bibinfo {author} {\bibfnamefont {R.~E.}\ \bibnamefont
  {Tribble}}, \bibinfo {author} {\bibfnamefont {K.~M.}\ \bibnamefont
  {Nollett}}, \bibinfo {author} {\bibfnamefont {F.~M.}\ \bibnamefont {Nunes}},
  \bibinfo {author} {\bibfnamefont {T.-S.}\ \bibnamefont {Park}}, \bibinfo
  {author} {\bibfnamefont {P.~D.}\ \bibnamefont {Parker}}, \bibinfo {author}
  {\bibfnamefont {R.}~\bibnamefont {Schiavilla}}, \bibinfo {author}
  {\bibfnamefont {E.~C.}\ \bibnamefont {Simpson}}, \bibinfo {author}
  {\bibfnamefont {C.}~\bibnamefont {Spitaleri}}, \bibinfo {author}
  {\bibfnamefont {F.}~\bibnamefont {Strieder}}, \bibinfo {author}
  {\bibfnamefont {H.-P.}\ \bibnamefont {Trautvetter}}, \bibinfo {author}
  {\bibfnamefont {K.}~\bibnamefont {Suemmerer}},\ and\ \bibinfo {author}
  {\bibfnamefont {S.}~\bibnamefont {Typel}},\ }\bibfield  {title} {\bibinfo
  {title} {Solar fusion cross sections. ii. the $pp$ chain and cno cycles},\
  }\href {https://doi.org/10.1103/RevModPhys.83.195} {\bibfield  {journal}
  {\bibinfo  {journal} {Rev. Mod. Phys.}\ }\textbf {\bibinfo {volume} {83}},\
  \bibinfo {pages} {195} (\bibinfo {year} {2011})}\BibitemShut {NoStop}%
\bibitem [{\citenamefont {Abe}\ \emph {et~al.}(2011)\citenamefont {Abe},
  \citenamefont {Hayato}, \citenamefont {Iida}, \citenamefont {Ikeda},
  \citenamefont {Ishihara}, \citenamefont {Iyogi}, \citenamefont {Kameda},
  \citenamefont {Kobayashi}, \citenamefont {Koshio}, \citenamefont {Kozuma},
  \citenamefont {Miura}, \citenamefont {Moriyama}, \citenamefont {Nakahata},
  \citenamefont {Nakayama}, \citenamefont {Obayashi}, \citenamefont {Ogawa},
  \citenamefont {Sekiya}, \citenamefont {Shiozawa}, \citenamefont {Suzuki},
  \citenamefont {Takeda}, \citenamefont {Takenaga}, \citenamefont {Ueno},
  \citenamefont {Ueshima}, \citenamefont {Watanabe}, \citenamefont {Yamada},
  \citenamefont {Yokozawa}, \citenamefont {Hazama}, \citenamefont {Kaji},
  \citenamefont {Kajita}, \citenamefont {Kaneyuki}, \citenamefont {McLachlan},
  \citenamefont {Okumura}, \citenamefont {Shimizu}, \citenamefont {Tanimoto},
  \citenamefont {Vagins}, \citenamefont {Labarga}, \citenamefont {Magro},
  \citenamefont {Dufour}, \citenamefont {Kearns}, \citenamefont {Litos},
  \citenamefont {Raaf}, \citenamefont {Stone}, \citenamefont {Sulak},
  \citenamefont {Wang}, \citenamefont {Goldhaber}, \citenamefont {Bays},
  \citenamefont {Casper}, \citenamefont {Cravens}, \citenamefont {Kropp},
  \citenamefont {Mine}, \citenamefont {Regis}, \citenamefont {Renshaw},
  \citenamefont {Smy}, \citenamefont {Sobel}, \citenamefont {Ganezer},
  \citenamefont {Hill}, \citenamefont {Keig}, \citenamefont {Jang},
  \citenamefont {Kim}, \citenamefont {Lim}, \citenamefont {Albert},
  \citenamefont {Wendell}, \citenamefont {Wongjirad}, \citenamefont
  {Scholberg}, \citenamefont {Walter}, \citenamefont {Ishizuka}, \citenamefont
  {Tasaka}, \citenamefont {Learned}, \citenamefont {Matsuno}, \citenamefont
  {Watanabe}, \citenamefont {Hasegawa}, \citenamefont {Ishida}, \citenamefont
  {Ishii}, \citenamefont {Kobayashi}, \citenamefont {Nakadaira}, \citenamefont
  {Nakamura}, \citenamefont {Nishikawa}, \citenamefont {Nishino}, \citenamefont
  {Oyama}, \citenamefont {Sakashita}, \citenamefont {Sekiguchi}, \citenamefont
  {Tsukamoto}, \citenamefont {Suzuki}, \citenamefont {Takeuchi}, \citenamefont
  {Minamino}, \citenamefont {Nakaya}, \citenamefont {Fukuda}, \citenamefont
  {Itow}, \citenamefont {Mitsuka}, \citenamefont {Tanaka}, \citenamefont
  {Jung}, \citenamefont {Lopez}, \citenamefont {McGrew}, \citenamefont {Terri},
  \citenamefont {Yanagisawa}, \citenamefont {Tamura}, \citenamefont {Ishino},
  \citenamefont {Kibayashi}, \citenamefont {Mino}, \citenamefont {Mori},
  \citenamefont {Sakuda}, \citenamefont {Toyota}, \citenamefont {Kuno},
  \citenamefont {Yoshida}, \citenamefont {Kim}, \citenamefont {Yang},
  \citenamefont {Ishizuka}, \citenamefont {Okazawa}, \citenamefont {Choi},
  \citenamefont {Nishijima}, \citenamefont {Yokosawa}, \citenamefont {Koshiba},
  \citenamefont {Totsuka}, \citenamefont {Yokoyama}, \citenamefont {Chen},
  \citenamefont {Heng}, \citenamefont {Yang}, \citenamefont {Zhang},
  \citenamefont {Kielczewska}, \citenamefont {Mijakowski}, \citenamefont
  {Connolly}, \citenamefont {Dziomba}, \citenamefont {Thrane},\ and\
  \citenamefont {Wilkes}}]{SuperK}%
  \BibitemOpen
  \bibfield  {author} {\bibinfo {author} {\bibfnamefont {K.}~\bibnamefont
  {Abe}}, \bibinfo {author} {\bibfnamefont {Y.}~\bibnamefont {Hayato}},
  \bibinfo {author} {\bibfnamefont {T.}~\bibnamefont {Iida}}, \bibinfo {author}
  {\bibfnamefont {M.}~\bibnamefont {Ikeda}}, \bibinfo {author} {\bibfnamefont
  {C.}~\bibnamefont {Ishihara}}, \bibinfo {author} {\bibfnamefont
  {K.}~\bibnamefont {Iyogi}}, \bibinfo {author} {\bibfnamefont
  {J.}~\bibnamefont {Kameda}}, \bibinfo {author} {\bibfnamefont
  {K.}~\bibnamefont {Kobayashi}}, \bibinfo {author} {\bibfnamefont
  {Y.}~\bibnamefont {Koshio}}, \bibinfo {author} {\bibfnamefont
  {Y.}~\bibnamefont {Kozuma}}, \bibinfo {author} {\bibfnamefont
  {M.}~\bibnamefont {Miura}}, \bibinfo {author} {\bibfnamefont
  {S.}~\bibnamefont {Moriyama}}, \bibinfo {author} {\bibfnamefont
  {M.}~\bibnamefont {Nakahata}}, \bibinfo {author} {\bibfnamefont
  {S.}~\bibnamefont {Nakayama}}, \bibinfo {author} {\bibfnamefont
  {Y.}~\bibnamefont {Obayashi}}, \bibinfo {author} {\bibfnamefont
  {H.}~\bibnamefont {Ogawa}}, \bibinfo {author} {\bibfnamefont
  {H.}~\bibnamefont {Sekiya}}, \bibinfo {author} {\bibfnamefont
  {M.}~\bibnamefont {Shiozawa}}, \bibinfo {author} {\bibfnamefont
  {Y.}~\bibnamefont {Suzuki}}, \bibinfo {author} {\bibfnamefont
  {A.}~\bibnamefont {Takeda}}, \bibinfo {author} {\bibfnamefont
  {Y.}~\bibnamefont {Takenaga}}, \bibinfo {author} {\bibfnamefont
  {K.}~\bibnamefont {Ueno}}, \bibinfo {author} {\bibfnamefont {K.}~\bibnamefont
  {Ueshima}}, \bibinfo {author} {\bibfnamefont {H.}~\bibnamefont {Watanabe}},
  \bibinfo {author} {\bibfnamefont {S.}~\bibnamefont {Yamada}}, \bibinfo
  {author} {\bibfnamefont {T.}~\bibnamefont {Yokozawa}}, \bibinfo {author}
  {\bibfnamefont {S.}~\bibnamefont {Hazama}}, \bibinfo {author} {\bibfnamefont
  {H.}~\bibnamefont {Kaji}}, \bibinfo {author} {\bibfnamefont {T.}~\bibnamefont
  {Kajita}}, \bibinfo {author} {\bibfnamefont {K.}~\bibnamefont {Kaneyuki}},
  \bibinfo {author} {\bibfnamefont {T.}~\bibnamefont {McLachlan}}, \bibinfo
  {author} {\bibfnamefont {K.}~\bibnamefont {Okumura}}, \bibinfo {author}
  {\bibfnamefont {Y.}~\bibnamefont {Shimizu}}, \bibinfo {author} {\bibfnamefont
  {N.}~\bibnamefont {Tanimoto}}, \bibinfo {author} {\bibfnamefont {M.~R.}\
  \bibnamefont {Vagins}}, \bibinfo {author} {\bibfnamefont {L.}~\bibnamefont
  {Labarga}}, \bibinfo {author} {\bibfnamefont {L.~M.}\ \bibnamefont {Magro}},
  \bibinfo {author} {\bibfnamefont {F.}~\bibnamefont {Dufour}}, \bibinfo
  {author} {\bibfnamefont {E.}~\bibnamefont {Kearns}}, \bibinfo {author}
  {\bibfnamefont {M.}~\bibnamefont {Litos}}, \bibinfo {author} {\bibfnamefont
  {J.~L.}\ \bibnamefont {Raaf}}, \bibinfo {author} {\bibfnamefont {J.~L.}\
  \bibnamefont {Stone}}, \bibinfo {author} {\bibfnamefont {L.~R.}\ \bibnamefont
  {Sulak}}, \bibinfo {author} {\bibfnamefont {W.}~\bibnamefont {Wang}},
  \bibinfo {author} {\bibfnamefont {M.}~\bibnamefont {Goldhaber}}, \bibinfo
  {author} {\bibfnamefont {K.}~\bibnamefont {Bays}}, \bibinfo {author}
  {\bibfnamefont {D.}~\bibnamefont {Casper}}, \bibinfo {author} {\bibfnamefont
  {J.~P.}\ \bibnamefont {Cravens}}, \bibinfo {author} {\bibfnamefont {W.~R.}\
  \bibnamefont {Kropp}}, \bibinfo {author} {\bibfnamefont {S.}~\bibnamefont
  {Mine}}, \bibinfo {author} {\bibfnamefont {C.}~\bibnamefont {Regis}},
  \bibinfo {author} {\bibfnamefont {A.}~\bibnamefont {Renshaw}}, \bibinfo
  {author} {\bibfnamefont {M.~B.}\ \bibnamefont {Smy}}, \bibinfo {author}
  {\bibfnamefont {H.~W.}\ \bibnamefont {Sobel}}, \bibinfo {author}
  {\bibfnamefont {K.~S.}\ \bibnamefont {Ganezer}}, \bibinfo {author}
  {\bibfnamefont {J.}~\bibnamefont {Hill}}, \bibinfo {author} {\bibfnamefont
  {W.~E.}\ \bibnamefont {Keig}}, \bibinfo {author} {\bibfnamefont {J.~S.}\
  \bibnamefont {Jang}}, \bibinfo {author} {\bibfnamefont {J.~Y.}\ \bibnamefont
  {Kim}}, \bibinfo {author} {\bibfnamefont {I.~T.}\ \bibnamefont {Lim}},
  \bibinfo {author} {\bibfnamefont {J.}~\bibnamefont {Albert}}, \bibinfo
  {author} {\bibfnamefont {R.}~\bibnamefont {Wendell}}, \bibinfo {author}
  {\bibfnamefont {T.}~\bibnamefont {Wongjirad}}, \bibinfo {author}
  {\bibfnamefont {K.}~\bibnamefont {Scholberg}}, \bibinfo {author}
  {\bibfnamefont {C.~W.}\ \bibnamefont {Walter}}, \bibinfo {author}
  {\bibfnamefont {T.}~\bibnamefont {Ishizuka}}, \bibinfo {author}
  {\bibfnamefont {S.}~\bibnamefont {Tasaka}}, \bibinfo {author} {\bibfnamefont
  {J.~G.}\ \bibnamefont {Learned}}, \bibinfo {author} {\bibfnamefont
  {S.}~\bibnamefont {Matsuno}}, \bibinfo {author} {\bibfnamefont
  {Y.}~\bibnamefont {Watanabe}}, \bibinfo {author} {\bibfnamefont
  {T.}~\bibnamefont {Hasegawa}}, \bibinfo {author} {\bibfnamefont
  {T.}~\bibnamefont {Ishida}}, \bibinfo {author} {\bibfnamefont
  {T.}~\bibnamefont {Ishii}}, \bibinfo {author} {\bibfnamefont
  {T.}~\bibnamefont {Kobayashi}}, \bibinfo {author} {\bibfnamefont
  {T.}~\bibnamefont {Nakadaira}}, \bibinfo {author} {\bibfnamefont
  {K.}~\bibnamefont {Nakamura}}, \bibinfo {author} {\bibfnamefont
  {K.}~\bibnamefont {Nishikawa}}, \bibinfo {author} {\bibfnamefont
  {H.}~\bibnamefont {Nishino}}, \bibinfo {author} {\bibfnamefont
  {Y.}~\bibnamefont {Oyama}}, \bibinfo {author} {\bibfnamefont
  {K.}~\bibnamefont {Sakashita}}, \bibinfo {author} {\bibfnamefont
  {T.}~\bibnamefont {Sekiguchi}}, \bibinfo {author} {\bibfnamefont
  {T.}~\bibnamefont {Tsukamoto}}, \bibinfo {author} {\bibfnamefont {A.~T.}\
  \bibnamefont {Suzuki}}, \bibinfo {author} {\bibfnamefont {Y.}~\bibnamefont
  {Takeuchi}}, \bibinfo {author} {\bibfnamefont {A.}~\bibnamefont {Minamino}},
  \bibinfo {author} {\bibfnamefont {T.}~\bibnamefont {Nakaya}}, \bibinfo
  {author} {\bibfnamefont {Y.}~\bibnamefont {Fukuda}}, \bibinfo {author}
  {\bibfnamefont {Y.}~\bibnamefont {Itow}}, \bibinfo {author} {\bibfnamefont
  {G.}~\bibnamefont {Mitsuka}}, \bibinfo {author} {\bibfnamefont
  {T.}~\bibnamefont {Tanaka}}, \bibinfo {author} {\bibfnamefont {C.~K.}\
  \bibnamefont {Jung}}, \bibinfo {author} {\bibfnamefont {G.}~\bibnamefont
  {Lopez}}, \bibinfo {author} {\bibfnamefont {C.}~\bibnamefont {McGrew}},
  \bibinfo {author} {\bibfnamefont {R.}~\bibnamefont {Terri}}, \bibinfo
  {author} {\bibfnamefont {C.}~\bibnamefont {Yanagisawa}}, \bibinfo {author}
  {\bibfnamefont {N.}~\bibnamefont {Tamura}}, \bibinfo {author} {\bibfnamefont
  {H.}~\bibnamefont {Ishino}}, \bibinfo {author} {\bibfnamefont
  {A.}~\bibnamefont {Kibayashi}}, \bibinfo {author} {\bibfnamefont
  {S.}~\bibnamefont {Mino}}, \bibinfo {author} {\bibfnamefont {T.}~\bibnamefont
  {Mori}}, \bibinfo {author} {\bibfnamefont {M.}~\bibnamefont {Sakuda}},
  \bibinfo {author} {\bibfnamefont {H.}~\bibnamefont {Toyota}}, \bibinfo
  {author} {\bibfnamefont {Y.}~\bibnamefont {Kuno}}, \bibinfo {author}
  {\bibfnamefont {M.}~\bibnamefont {Yoshida}}, \bibinfo {author} {\bibfnamefont
  {S.~B.}\ \bibnamefont {Kim}}, \bibinfo {author} {\bibfnamefont {B.~S.}\
  \bibnamefont {Yang}}, \bibinfo {author} {\bibfnamefont {T.}~\bibnamefont
  {Ishizuka}}, \bibinfo {author} {\bibfnamefont {H.}~\bibnamefont {Okazawa}},
  \bibinfo {author} {\bibfnamefont {Y.}~\bibnamefont {Choi}}, \bibinfo {author}
  {\bibfnamefont {K.}~\bibnamefont {Nishijima}}, \bibinfo {author}
  {\bibfnamefont {Y.}~\bibnamefont {Yokosawa}}, \bibinfo {author}
  {\bibfnamefont {M.}~\bibnamefont {Koshiba}}, \bibinfo {author} {\bibfnamefont
  {Y.}~\bibnamefont {Totsuka}}, \bibinfo {author} {\bibfnamefont
  {M.}~\bibnamefont {Yokoyama}}, \bibinfo {author} {\bibfnamefont
  {S.}~\bibnamefont {Chen}}, \bibinfo {author} {\bibfnamefont {Y.}~\bibnamefont
  {Heng}}, \bibinfo {author} {\bibfnamefont {Z.}~\bibnamefont {Yang}}, \bibinfo
  {author} {\bibfnamefont {H.}~\bibnamefont {Zhang}}, \bibinfo {author}
  {\bibfnamefont {D.}~\bibnamefont {Kielczewska}}, \bibinfo {author}
  {\bibfnamefont {P.}~\bibnamefont {Mijakowski}}, \bibinfo {author}
  {\bibfnamefont {K.}~\bibnamefont {Connolly}}, \bibinfo {author}
  {\bibfnamefont {M.}~\bibnamefont {Dziomba}}, \bibinfo {author} {\bibfnamefont
  {E.}~\bibnamefont {Thrane}},\ and\ \bibinfo {author} {\bibfnamefont {R.~J.}\
  \bibnamefont {Wilkes}} (\bibinfo {collaboration} {Super-Kamiokande
  Collaboration}),\ }\bibfield  {title} {\bibinfo {title} {Solar neutrino
  results in super-kamiokande-iii},\ }\href
  {https://doi.org/10.1103/PhysRevD.83.052010} {\bibfield  {journal} {\bibinfo
  {journal} {Phys. Rev. D}\ }\textbf {\bibinfo {volume} {83}},\ \bibinfo
  {pages} {052010} (\bibinfo {year} {2011})}\BibitemShut {NoStop}%
\bibitem [{\citenamefont {Ahmad}\ \emph {et~al.}(2002)\citenamefont {Ahmad},
  \citenamefont {Allen}, \citenamefont {Andersen}, \citenamefont {D.Anglin},
  \citenamefont {Barton}, \citenamefont {Beier}, \citenamefont {Bercovitch},
  \citenamefont {Bigu}, \citenamefont {Biller}, \citenamefont {Black},
  \citenamefont {Blevis}, \citenamefont {Boardman}, \citenamefont {Boger},
  \citenamefont {Bonvin}, \citenamefont {Boulay}, \citenamefont {Bowler},
  \citenamefont {Bowles}, \citenamefont {Brice}, \citenamefont {Browne},
  \citenamefont {Bullard}, \citenamefont {B\"uhler}, \citenamefont {Cameron},
  \citenamefont {Chan}, \citenamefont {Chen}, \citenamefont {Chen},
  \citenamefont {Chen}, \citenamefont {Cleveland}, \citenamefont {Clifford},
  \citenamefont {Cowan}, \citenamefont {Cowen}, \citenamefont {Cox},
  \citenamefont {Dai}, \citenamefont {Dalnoki-Veress}, \citenamefont
  {Davidson}, \citenamefont {Doe}, \citenamefont {Doucas}, \citenamefont
  {Dragowsky}, \citenamefont {Duba}, \citenamefont {Duncan}, \citenamefont
  {Dunford}, \citenamefont {Dunmore}, \citenamefont {Earle}, \citenamefont
  {Elliott}, \citenamefont {Evans}, \citenamefont {Ewan}, \citenamefont
  {Farine}, \citenamefont {Fergani}, \citenamefont {Ferraris}, \citenamefont
  {Ford}, \citenamefont {Formaggio}, \citenamefont {Fowler}, \citenamefont
  {Frame}, \citenamefont {Frank}, \citenamefont {Frati}, \citenamefont
  {Gagnon}, \citenamefont {Germani}, \citenamefont {Gil}, \citenamefont
  {Graham}, \citenamefont {Grant}, \citenamefont {Hahn}, \citenamefont
  {Hallin}, \citenamefont {Hallman}, \citenamefont {Hamer}, \citenamefont
  {Hamian}, \citenamefont {Handler}, \citenamefont {Haq}, \citenamefont
  {Hargrove}, \citenamefont {Harvey}, \citenamefont {Hazama}, \citenamefont
  {Heeger}, \citenamefont {Heintzelman}, \citenamefont {Heise}, \citenamefont
  {Helmer}, \citenamefont {Hepburn}, \citenamefont {Heron}, \citenamefont
  {Hewett}, \citenamefont {Hime}, \citenamefont {Howe}, \citenamefont {Hykawy},
  \citenamefont {Isaac}, \citenamefont {Jagam}, \citenamefont {Jelley},
  \citenamefont {Jillings}, \citenamefont {Jonkmans}, \citenamefont {Kazkaz},
  \citenamefont {Keener}, \citenamefont {Klein}, \citenamefont {Knox},
  \citenamefont {Komar}, \citenamefont {Kouzes}, \citenamefont {Kutter},
  \citenamefont {Kyba}, \citenamefont {Law}, \citenamefont {Lawson},
  \citenamefont {Lay}, \citenamefont {Lee}, \citenamefont {Lesko},
  \citenamefont {Leslie}, \citenamefont {Levine}, \citenamefont {Locke},
  \citenamefont {Luoma}, \citenamefont {Lyon}, \citenamefont {Majerus},
  \citenamefont {Mak}, \citenamefont {Maneira}, \citenamefont {Manor},
  \citenamefont {Marino}, \citenamefont {McCauley}, \citenamefont {McDonald},
  \citenamefont {McDonald}, \citenamefont {McFarlane}, \citenamefont
  {McGregor}, \citenamefont {Meijer~Drees}, \citenamefont {Mifflin},
  \citenamefont {Miller}, \citenamefont {Milton}, \citenamefont {Moffat},
  \citenamefont {Moorhead}, \citenamefont {Nally}, \citenamefont {Neubauer},
  \citenamefont {Newcomer}, \citenamefont {Ng}, \citenamefont {Noble},
  \citenamefont {Norman}, \citenamefont {Novikov}, \citenamefont {O'Neill},
  \citenamefont {Okada}, \citenamefont {Ollerhead}, \citenamefont {Omori},
  \citenamefont {Orrell}, \citenamefont {Oser}, \citenamefont {Poon},
  \citenamefont {Radcliffe}, \citenamefont {Roberge}, \citenamefont
  {Robertson}, \citenamefont {Robertson}, \citenamefont {Rosendahl},
  \citenamefont {Rowley}, \citenamefont {Rusu}, \citenamefont {Saettler},
  \citenamefont {Schaffer}, \citenamefont {Schwendener}, \citenamefont
  {Sch\"ulke}, \citenamefont {Seifert}, \citenamefont {Shatkay}, \citenamefont
  {Simpson}, \citenamefont {Sims}, \citenamefont {Sinclair}, \citenamefont
  {Skensved}, \citenamefont {Smith}, \citenamefont {Smith}, \citenamefont
  {Spreitzer}, \citenamefont {Starinsky}, \citenamefont {Steiger},
  \citenamefont {Stokstad}, \citenamefont {Stonehill}, \citenamefont {Storey},
  \citenamefont {Sur}, \citenamefont {Tafirout}, \citenamefont {Tagg},
  \citenamefont {Tanner}, \citenamefont {Taplin}, \citenamefont {Thorman},
  \citenamefont {Thornewell}, \citenamefont {Trent}, \citenamefont
  {Tserkovnyak}, \citenamefont {Van~Berg}, \citenamefont {Van~de Water},
  \citenamefont {Virtue}, \citenamefont {Waltham}, \citenamefont {Wang},
  \citenamefont {Wark}, \citenamefont {West}, \citenamefont {Wilhelmy},
  \citenamefont {Wilkerson}, \citenamefont {Wilson}, \citenamefont {Wittich},
  \citenamefont {Wouters},\ and\ \citenamefont {Yeh}}]{SNOcollaboration}%
  \BibitemOpen
  \bibfield  {author} {\bibinfo {author} {\bibfnamefont {Q.~R.}\ \bibnamefont
  {Ahmad}}, \bibinfo {author} {\bibfnamefont {R.~C.}\ \bibnamefont {Allen}},
  \bibinfo {author} {\bibfnamefont {T.~C.}\ \bibnamefont {Andersen}}, \bibinfo
  {author} {\bibfnamefont {J.}~\bibnamefont {D.Anglin}}, \bibinfo {author}
  {\bibfnamefont {J.~C.}\ \bibnamefont {Barton}}, \bibinfo {author}
  {\bibfnamefont {E.~W.}\ \bibnamefont {Beier}}, \bibinfo {author}
  {\bibfnamefont {M.}~\bibnamefont {Bercovitch}}, \bibinfo {author}
  {\bibfnamefont {J.}~\bibnamefont {Bigu}}, \bibinfo {author} {\bibfnamefont
  {S.~D.}\ \bibnamefont {Biller}}, \bibinfo {author} {\bibfnamefont {R.~A.}\
  \bibnamefont {Black}}, \bibinfo {author} {\bibfnamefont {I.}~\bibnamefont
  {Blevis}}, \bibinfo {author} {\bibfnamefont {R.~J.}\ \bibnamefont
  {Boardman}}, \bibinfo {author} {\bibfnamefont {J.}~\bibnamefont {Boger}},
  \bibinfo {author} {\bibfnamefont {E.}~\bibnamefont {Bonvin}}, \bibinfo
  {author} {\bibfnamefont {M.~G.}\ \bibnamefont {Boulay}}, \bibinfo {author}
  {\bibfnamefont {M.~G.}\ \bibnamefont {Bowler}}, \bibinfo {author}
  {\bibfnamefont {T.~J.}\ \bibnamefont {Bowles}}, \bibinfo {author}
  {\bibfnamefont {S.~J.}\ \bibnamefont {Brice}}, \bibinfo {author}
  {\bibfnamefont {M.~C.}\ \bibnamefont {Browne}}, \bibinfo {author}
  {\bibfnamefont {T.~V.}\ \bibnamefont {Bullard}}, \bibinfo {author}
  {\bibfnamefont {G.}~\bibnamefont {B\"uhler}}, \bibinfo {author}
  {\bibfnamefont {J.}~\bibnamefont {Cameron}}, \bibinfo {author} {\bibfnamefont
  {Y.~D.}\ \bibnamefont {Chan}}, \bibinfo {author} {\bibfnamefont {H.~H.}\
  \bibnamefont {Chen}}, \bibinfo {author} {\bibfnamefont {M.}~\bibnamefont
  {Chen}}, \bibinfo {author} {\bibfnamefont {X.}~\bibnamefont {Chen}}, \bibinfo
  {author} {\bibfnamefont {B.~T.}\ \bibnamefont {Cleveland}}, \bibinfo {author}
  {\bibfnamefont {E.~T.~H.}\ \bibnamefont {Clifford}}, \bibinfo {author}
  {\bibfnamefont {J.~H.~M.}\ \bibnamefont {Cowan}}, \bibinfo {author}
  {\bibfnamefont {D.~F.}\ \bibnamefont {Cowen}}, \bibinfo {author}
  {\bibfnamefont {G.~A.}\ \bibnamefont {Cox}}, \bibinfo {author} {\bibfnamefont
  {X.}~\bibnamefont {Dai}}, \bibinfo {author} {\bibfnamefont {F.}~\bibnamefont
  {Dalnoki-Veress}}, \bibinfo {author} {\bibfnamefont {W.~F.}\ \bibnamefont
  {Davidson}}, \bibinfo {author} {\bibfnamefont {P.~J.}\ \bibnamefont {Doe}},
  \bibinfo {author} {\bibfnamefont {G.}~\bibnamefont {Doucas}}, \bibinfo
  {author} {\bibfnamefont {M.~R.}\ \bibnamefont {Dragowsky}}, \bibinfo {author}
  {\bibfnamefont {C.~A.}\ \bibnamefont {Duba}}, \bibinfo {author}
  {\bibfnamefont {F.~A.}\ \bibnamefont {Duncan}}, \bibinfo {author}
  {\bibfnamefont {M.}~\bibnamefont {Dunford}}, \bibinfo {author} {\bibfnamefont
  {J.~A.}\ \bibnamefont {Dunmore}}, \bibinfo {author} {\bibfnamefont {E.~D.}\
  \bibnamefont {Earle}}, \bibinfo {author} {\bibfnamefont {S.~R.}\ \bibnamefont
  {Elliott}}, \bibinfo {author} {\bibfnamefont {H.~C.}\ \bibnamefont {Evans}},
  \bibinfo {author} {\bibfnamefont {G.~T.}\ \bibnamefont {Ewan}}, \bibinfo
  {author} {\bibfnamefont {J.}~\bibnamefont {Farine}}, \bibinfo {author}
  {\bibfnamefont {H.}~\bibnamefont {Fergani}}, \bibinfo {author} {\bibfnamefont
  {A.~P.}\ \bibnamefont {Ferraris}}, \bibinfo {author} {\bibfnamefont {R.~J.}\
  \bibnamefont {Ford}}, \bibinfo {author} {\bibfnamefont {J.~A.}\ \bibnamefont
  {Formaggio}}, \bibinfo {author} {\bibfnamefont {M.~M.}\ \bibnamefont
  {Fowler}}, \bibinfo {author} {\bibfnamefont {K.}~\bibnamefont {Frame}},
  \bibinfo {author} {\bibfnamefont {E.~D.}\ \bibnamefont {Frank}}, \bibinfo
  {author} {\bibfnamefont {W.}~\bibnamefont {Frati}}, \bibinfo {author}
  {\bibfnamefont {N.}~\bibnamefont {Gagnon}}, \bibinfo {author} {\bibfnamefont
  {J.~V.}\ \bibnamefont {Germani}}, \bibinfo {author} {\bibfnamefont
  {S.}~\bibnamefont {Gil}}, \bibinfo {author} {\bibfnamefont {K.}~\bibnamefont
  {Graham}}, \bibinfo {author} {\bibfnamefont {D.~R.}\ \bibnamefont {Grant}},
  \bibinfo {author} {\bibfnamefont {R.~L.}\ \bibnamefont {Hahn}}, \bibinfo
  {author} {\bibfnamefont {A.~L.}\ \bibnamefont {Hallin}}, \bibinfo {author}
  {\bibfnamefont {E.~D.}\ \bibnamefont {Hallman}}, \bibinfo {author}
  {\bibfnamefont {A.~S.}\ \bibnamefont {Hamer}}, \bibinfo {author}
  {\bibfnamefont {A.~A.}\ \bibnamefont {Hamian}}, \bibinfo {author}
  {\bibfnamefont {W.~B.}\ \bibnamefont {Handler}}, \bibinfo {author}
  {\bibfnamefont {R.~U.}\ \bibnamefont {Haq}}, \bibinfo {author} {\bibfnamefont
  {C.~K.}\ \bibnamefont {Hargrove}}, \bibinfo {author} {\bibfnamefont {P.~J.}\
  \bibnamefont {Harvey}}, \bibinfo {author} {\bibfnamefont {R.}~\bibnamefont
  {Hazama}}, \bibinfo {author} {\bibfnamefont {K.~M.}\ \bibnamefont {Heeger}},
  \bibinfo {author} {\bibfnamefont {W.~J.}\ \bibnamefont {Heintzelman}},
  \bibinfo {author} {\bibfnamefont {J.}~\bibnamefont {Heise}}, \bibinfo
  {author} {\bibfnamefont {R.~L.}\ \bibnamefont {Helmer}}, \bibinfo {author}
  {\bibfnamefont {J.~D.}\ \bibnamefont {Hepburn}}, \bibinfo {author}
  {\bibfnamefont {H.}~\bibnamefont {Heron}}, \bibinfo {author} {\bibfnamefont
  {J.}~\bibnamefont {Hewett}}, \bibinfo {author} {\bibfnamefont
  {A.}~\bibnamefont {Hime}}, \bibinfo {author} {\bibfnamefont {M.}~\bibnamefont
  {Howe}}, \bibinfo {author} {\bibfnamefont {J.~G.}\ \bibnamefont {Hykawy}},
  \bibinfo {author} {\bibfnamefont {M.~C.~P.}\ \bibnamefont {Isaac}}, \bibinfo
  {author} {\bibfnamefont {P.}~\bibnamefont {Jagam}}, \bibinfo {author}
  {\bibfnamefont {N.~A.}\ \bibnamefont {Jelley}}, \bibinfo {author}
  {\bibfnamefont {C.}~\bibnamefont {Jillings}}, \bibinfo {author}
  {\bibfnamefont {G.}~\bibnamefont {Jonkmans}}, \bibinfo {author}
  {\bibfnamefont {K.}~\bibnamefont {Kazkaz}}, \bibinfo {author} {\bibfnamefont
  {P.~T.}\ \bibnamefont {Keener}}, \bibinfo {author} {\bibfnamefont {J.~R.}\
  \bibnamefont {Klein}}, \bibinfo {author} {\bibfnamefont {A.~B.}\ \bibnamefont
  {Knox}}, \bibinfo {author} {\bibfnamefont {R.~J.}\ \bibnamefont {Komar}},
  \bibinfo {author} {\bibfnamefont {R.}~\bibnamefont {Kouzes}}, \bibinfo
  {author} {\bibfnamefont {T.}~\bibnamefont {Kutter}}, \bibinfo {author}
  {\bibfnamefont {C.~C.~M.}\ \bibnamefont {Kyba}}, \bibinfo {author}
  {\bibfnamefont {J.}~\bibnamefont {Law}}, \bibinfo {author} {\bibfnamefont
  {I.~T.}\ \bibnamefont {Lawson}}, \bibinfo {author} {\bibfnamefont
  {M.}~\bibnamefont {Lay}}, \bibinfo {author} {\bibfnamefont {H.~W.}\
  \bibnamefont {Lee}}, \bibinfo {author} {\bibfnamefont {K.~T.}\ \bibnamefont
  {Lesko}}, \bibinfo {author} {\bibfnamefont {J.~R.}\ \bibnamefont {Leslie}},
  \bibinfo {author} {\bibfnamefont {I.}~\bibnamefont {Levine}}, \bibinfo
  {author} {\bibfnamefont {W.}~\bibnamefont {Locke}}, \bibinfo {author}
  {\bibfnamefont {S.}~\bibnamefont {Luoma}}, \bibinfo {author} {\bibfnamefont
  {J.}~\bibnamefont {Lyon}}, \bibinfo {author} {\bibfnamefont {S.}~\bibnamefont
  {Majerus}}, \bibinfo {author} {\bibfnamefont {H.~B.}\ \bibnamefont {Mak}},
  \bibinfo {author} {\bibfnamefont {J.}~\bibnamefont {Maneira}}, \bibinfo
  {author} {\bibfnamefont {J.}~\bibnamefont {Manor}}, \bibinfo {author}
  {\bibfnamefont {A.~D.}\ \bibnamefont {Marino}}, \bibinfo {author}
  {\bibfnamefont {N.}~\bibnamefont {McCauley}}, \bibinfo {author}
  {\bibfnamefont {A.~B.}\ \bibnamefont {McDonald}}, \bibinfo {author}
  {\bibfnamefont {D.~S.}\ \bibnamefont {McDonald}}, \bibinfo {author}
  {\bibfnamefont {K.}~\bibnamefont {McFarlane}}, \bibinfo {author}
  {\bibfnamefont {G.}~\bibnamefont {McGregor}}, \bibinfo {author}
  {\bibfnamefont {R.}~\bibnamefont {Meijer~Drees}}, \bibinfo {author}
  {\bibfnamefont {C.}~\bibnamefont {Mifflin}}, \bibinfo {author} {\bibfnamefont
  {G.~G.}\ \bibnamefont {Miller}}, \bibinfo {author} {\bibfnamefont
  {G.}~\bibnamefont {Milton}}, \bibinfo {author} {\bibfnamefont {B.~A.}\
  \bibnamefont {Moffat}}, \bibinfo {author} {\bibfnamefont {M.}~\bibnamefont
  {Moorhead}}, \bibinfo {author} {\bibfnamefont {C.~W.}\ \bibnamefont {Nally}},
  \bibinfo {author} {\bibfnamefont {M.~S.}\ \bibnamefont {Neubauer}}, \bibinfo
  {author} {\bibfnamefont {F.~M.}\ \bibnamefont {Newcomer}}, \bibinfo {author}
  {\bibfnamefont {H.~S.}\ \bibnamefont {Ng}}, \bibinfo {author} {\bibfnamefont
  {A.~J.}\ \bibnamefont {Noble}}, \bibinfo {author} {\bibfnamefont {E.~B.}\
  \bibnamefont {Norman}}, \bibinfo {author} {\bibfnamefont {V.~M.}\
  \bibnamefont {Novikov}}, \bibinfo {author} {\bibfnamefont {M.}~\bibnamefont
  {O'Neill}}, \bibinfo {author} {\bibfnamefont {C.~E.}\ \bibnamefont {Okada}},
  \bibinfo {author} {\bibfnamefont {R.~W.}\ \bibnamefont {Ollerhead}}, \bibinfo
  {author} {\bibfnamefont {M.}~\bibnamefont {Omori}}, \bibinfo {author}
  {\bibfnamefont {J.~L.}\ \bibnamefont {Orrell}}, \bibinfo {author}
  {\bibfnamefont {S.~M.}\ \bibnamefont {Oser}}, \bibinfo {author}
  {\bibfnamefont {A.~W.~P.}\ \bibnamefont {Poon}}, \bibinfo {author}
  {\bibfnamefont {T.~J.}\ \bibnamefont {Radcliffe}}, \bibinfo {author}
  {\bibfnamefont {A.}~\bibnamefont {Roberge}}, \bibinfo {author} {\bibfnamefont
  {B.~C.}\ \bibnamefont {Robertson}}, \bibinfo {author} {\bibfnamefont
  {R.~G.~H.}\ \bibnamefont {Robertson}}, \bibinfo {author} {\bibfnamefont
  {S.~S.~E.}\ \bibnamefont {Rosendahl}}, \bibinfo {author} {\bibfnamefont
  {J.~K.}\ \bibnamefont {Rowley}}, \bibinfo {author} {\bibfnamefont {V.~L.}\
  \bibnamefont {Rusu}}, \bibinfo {author} {\bibfnamefont {E.}~\bibnamefont
  {Saettler}}, \bibinfo {author} {\bibfnamefont {K.~K.}\ \bibnamefont
  {Schaffer}}, \bibinfo {author} {\bibfnamefont {M.~H.}\ \bibnamefont
  {Schwendener}}, \bibinfo {author} {\bibfnamefont {A.}~\bibnamefont
  {Sch\"ulke}}, \bibinfo {author} {\bibfnamefont {H.}~\bibnamefont {Seifert}},
  \bibinfo {author} {\bibfnamefont {M.}~\bibnamefont {Shatkay}}, \bibinfo
  {author} {\bibfnamefont {J.~J.}\ \bibnamefont {Simpson}}, \bibinfo {author}
  {\bibfnamefont {C.~J.}\ \bibnamefont {Sims}}, \bibinfo {author}
  {\bibfnamefont {D.}~\bibnamefont {Sinclair}}, \bibinfo {author}
  {\bibfnamefont {P.}~\bibnamefont {Skensved}}, \bibinfo {author}
  {\bibfnamefont {A.~R.}\ \bibnamefont {Smith}}, \bibinfo {author}
  {\bibfnamefont {M.~W.~E.}\ \bibnamefont {Smith}}, \bibinfo {author}
  {\bibfnamefont {T.}~\bibnamefont {Spreitzer}}, \bibinfo {author}
  {\bibfnamefont {N.}~\bibnamefont {Starinsky}}, \bibinfo {author}
  {\bibfnamefont {T.~D.}\ \bibnamefont {Steiger}}, \bibinfo {author}
  {\bibfnamefont {R.~G.}\ \bibnamefont {Stokstad}}, \bibinfo {author}
  {\bibfnamefont {L.~C.}\ \bibnamefont {Stonehill}}, \bibinfo {author}
  {\bibfnamefont {R.~S.}\ \bibnamefont {Storey}}, \bibinfo {author}
  {\bibfnamefont {B.}~\bibnamefont {Sur}}, \bibinfo {author} {\bibfnamefont
  {R.}~\bibnamefont {Tafirout}}, \bibinfo {author} {\bibfnamefont
  {N.}~\bibnamefont {Tagg}}, \bibinfo {author} {\bibfnamefont {N.~W.}\
  \bibnamefont {Tanner}}, \bibinfo {author} {\bibfnamefont {R.~K.}\
  \bibnamefont {Taplin}}, \bibinfo {author} {\bibfnamefont {M.}~\bibnamefont
  {Thorman}}, \bibinfo {author} {\bibfnamefont {P.~M.}\ \bibnamefont
  {Thornewell}}, \bibinfo {author} {\bibfnamefont {P.~T.}\ \bibnamefont
  {Trent}}, \bibinfo {author} {\bibfnamefont {Y.~I.}\ \bibnamefont
  {Tserkovnyak}}, \bibinfo {author} {\bibfnamefont {R.}~\bibnamefont
  {Van~Berg}}, \bibinfo {author} {\bibfnamefont {R.~G.}\ \bibnamefont {Van~de
  Water}}, \bibinfo {author} {\bibfnamefont {C.~J.}\ \bibnamefont {Virtue}},
  \bibinfo {author} {\bibfnamefont {C.~E.}\ \bibnamefont {Waltham}}, \bibinfo
  {author} {\bibfnamefont {J.-X.}\ \bibnamefont {Wang}}, \bibinfo {author}
  {\bibfnamefont {D.~L.}\ \bibnamefont {Wark}}, \bibinfo {author}
  {\bibfnamefont {N.}~\bibnamefont {West}}, \bibinfo {author} {\bibfnamefont
  {J.~B.}\ \bibnamefont {Wilhelmy}}, \bibinfo {author} {\bibfnamefont {J.~F.}\
  \bibnamefont {Wilkerson}}, \bibinfo {author} {\bibfnamefont {J.~R.}\
  \bibnamefont {Wilson}}, \bibinfo {author} {\bibfnamefont {P.}~\bibnamefont
  {Wittich}}, \bibinfo {author} {\bibfnamefont {J.~M.}\ \bibnamefont
  {Wouters}},\ and\ \bibinfo {author} {\bibfnamefont {M.}~\bibnamefont {Yeh}}
  (\bibinfo {collaboration} {SNO Collaboration}),\ }\bibfield  {title}
  {\bibinfo {title} {Direct evidence for neutrino flavor transformation from
  neutral-current interactions in the sudbury neutrino observatory},\ }\href
  {https://doi.org/10.1103/PhysRevLett.89.011301} {\bibfield  {journal}
  {\bibinfo  {journal} {Phys. Rev. Lett.}\ }\textbf {\bibinfo {volume} {89}},\
  \bibinfo {pages} {011301} (\bibinfo {year} {2002})}\BibitemShut {NoStop}%
\bibitem [{\citenamefont {Filippone}\ \emph {et~al.}(1983)\citenamefont
  {Filippone}, \citenamefont {Elwyn}, \citenamefont {Davids},\ and\
  \citenamefont {Koetke}}]{Filippone1983}%
  \BibitemOpen
  \bibfield  {author} {\bibinfo {author} {\bibfnamefont {B.~W.}\ \bibnamefont
  {Filippone}}, \bibinfo {author} {\bibfnamefont {A.~J.}\ \bibnamefont
  {Elwyn}}, \bibinfo {author} {\bibfnamefont {C.~N.}\ \bibnamefont {Davids}},\
  and\ \bibinfo {author} {\bibfnamefont {D.~D.}\ \bibnamefont {Koetke}},\
  }\bibfield  {title} {\bibinfo {title} {Measurement of the $^7\mathrm{Be}(p,
  \ensuremath{\gamma})^{8}\mathrm{B}$ reaction cross section at low energies},\
  }\href {https://doi.org/10.1103/PhysRevLett.50.412} {\bibfield  {journal}
  {\bibinfo  {journal} {Phys. Rev. Lett.}\ }\textbf {\bibinfo {volume} {50}},\
  \bibinfo {pages} {412} (\bibinfo {year} {1983})}\BibitemShut {NoStop}%
\bibitem [{\citenamefont {Baur}\ \emph {et~al.}(1986)\citenamefont {Baur},
  \citenamefont {Bertulani},\ and\ \citenamefont {Rebel}}]{Baur1986}%
  \BibitemOpen
  \bibfield  {author} {\bibinfo {author} {\bibfnamefont {G.}~\bibnamefont
  {Baur}}, \bibinfo {author} {\bibfnamefont {C.}~\bibnamefont {Bertulani}},\
  and\ \bibinfo {author} {\bibfnamefont {H.}~\bibnamefont {Rebel}},\ }\bibfield
   {title} {\bibinfo {title} {Coulomb dissociation as a source of information
  on radiative capture processes of astrophysical interest},\ }\href
  {https://doi.org/https://doi.org/10.1016/0375-9474(86)90290-3} {\bibfield
  {journal} {\bibinfo  {journal} {Nuclear Physics A}\ }\textbf {\bibinfo
  {volume} {458}},\ \bibinfo {pages} {188} (\bibinfo {year}
  {1986})}\BibitemShut {NoStop}%
\bibitem [{\citenamefont {Kikuchi}\ \emph {et~al.}(1998)\citenamefont
  {Kikuchi}, \citenamefont {Motobayashi}, \citenamefont {Iwasa}, \citenamefont
  {Ando}, \citenamefont {Kurokawa}, \citenamefont {Moriya}, \citenamefont
  {Murakami}, \citenamefont {Nishio}, \citenamefont {Ruan~(Gen)}, \citenamefont
  {Shirato}, \citenamefont {Shimoura}, \citenamefont {Uchibori}, \citenamefont
  {Yanagisawa}, \citenamefont {Kubo}, \citenamefont {Sakurai}, \citenamefont
  {Teranishi}, \citenamefont {Watanabe}, \citenamefont {Ishihara},
  \citenamefont {Hirai}, \citenamefont {Nakamura}, \citenamefont {Kubono},
  \citenamefont {Gai}, \citenamefont {France~III}, \citenamefont {Hahn},
  \citenamefont {Delbar}, \citenamefont {Lipnik},\ and\ \citenamefont
  {Michotte}}]{Kikuchi1998}%
  \BibitemOpen
  \bibfield  {author} {\bibinfo {author} {\bibfnamefont {T.}~\bibnamefont
  {Kikuchi}}, \bibinfo {author} {\bibfnamefont {T.}~\bibnamefont
  {Motobayashi}}, \bibinfo {author} {\bibfnamefont {N.}~\bibnamefont {Iwasa}},
  \bibinfo {author} {\bibfnamefont {Y.}~\bibnamefont {Ando}}, \bibinfo {author}
  {\bibfnamefont {M.}~\bibnamefont {Kurokawa}}, \bibinfo {author}
  {\bibfnamefont {S.}~\bibnamefont {Moriya}}, \bibinfo {author} {\bibfnamefont
  {H.}~\bibnamefont {Murakami}}, \bibinfo {author} {\bibfnamefont
  {T.}~\bibnamefont {Nishio}}, \bibinfo {author} {\bibfnamefont
  {J.}~\bibnamefont {Ruan~(Gen)}}, \bibinfo {author} {\bibfnamefont
  {S.}~\bibnamefont {Shirato}}, \bibinfo {author} {\bibfnamefont
  {S.}~\bibnamefont {Shimoura}}, \bibinfo {author} {\bibfnamefont
  {T.}~\bibnamefont {Uchibori}}, \bibinfo {author} {\bibfnamefont
  {Y.}~\bibnamefont {Yanagisawa}}, \bibinfo {author} {\bibfnamefont
  {T.}~\bibnamefont {Kubo}}, \bibinfo {author} {\bibfnamefont {H.}~\bibnamefont
  {Sakurai}}, \bibinfo {author} {\bibfnamefont {T.}~\bibnamefont {Teranishi}},
  \bibinfo {author} {\bibfnamefont {Y.}~\bibnamefont {Watanabe}}, \bibinfo
  {author} {\bibfnamefont {M.}~\bibnamefont {Ishihara}}, \bibinfo {author}
  {\bibfnamefont {M.}~\bibnamefont {Hirai}}, \bibinfo {author} {\bibfnamefont
  {T.}~\bibnamefont {Nakamura}}, \bibinfo {author} {\bibfnamefont
  {S.}~\bibnamefont {Kubono}}, \bibinfo {author} {\bibfnamefont
  {M.}~\bibnamefont {Gai}}, \bibinfo {author} {\bibfnamefont {R.~H.}\
  \bibnamefont {France~III}}, \bibinfo {author} {\bibfnamefont {K.~I.}\
  \bibnamefont {Hahn}}, \bibinfo {author} {\bibfnamefont {T.}~\bibnamefont
  {Delbar}}, \bibinfo {author} {\bibfnamefont {P.}~\bibnamefont {Lipnik}},\
  and\ \bibinfo {author} {\bibfnamefont {C.}~\bibnamefont {Michotte}},\
  }\bibfield  {title} {\bibinfo {title} {Further measurement of the
  7be(p,\ensuremath{\gamma})8b cross section at low energies with the coulomb
  dissociation of 8b},\ }\href {https://doi.org/10.1007/s100500050170}
  {\bibfield  {journal} {\bibinfo  {journal} {The European Physical Journal A -
  Hadrons and Nuclei}\ }\textbf {\bibinfo {volume} {3}},\ \bibinfo {pages}
  {213} (\bibinfo {year} {1998})}\BibitemShut {NoStop}%
\bibitem [{\citenamefont {Iwasa}\ \emph {et~al.}(1999)\citenamefont {Iwasa},
  \citenamefont {Bou\'e}, \citenamefont {Sur\'owka}, \citenamefont
  {S\"ummerer}, \citenamefont {Baumann}, \citenamefont {Blank}, \citenamefont
  {Czajkowski}, \citenamefont {F\"orster}, \citenamefont {Gai}, \citenamefont
  {Geissel}, \citenamefont {Grosse}, \citenamefont {Hellstr\"om}, \citenamefont
  {Koczon}, \citenamefont {Kohlmeyer}, \citenamefont {Kulessa}, \citenamefont
  {Laue}, \citenamefont {Marchand}, \citenamefont {Motobayashi}, \citenamefont
  {Oeschler}, \citenamefont {Ozawa}, \citenamefont {Pravikoff}, \citenamefont
  {Schwab}, \citenamefont {Schwab}, \citenamefont {Senger}, \citenamefont
  {Speer}, \citenamefont {Sturm}, \citenamefont {Surowiec}, \citenamefont
  {Teranishi}, \citenamefont {Uhlig}, \citenamefont {Wagner}, \citenamefont
  {Walus},\ and\ \citenamefont {Bertulani}}]{Iwasa1999}%
  \BibitemOpen
  \bibfield  {author} {\bibinfo {author} {\bibfnamefont {N.}~\bibnamefont
  {Iwasa}}, \bibinfo {author} {\bibfnamefont {F.}~\bibnamefont {Bou\'e}},
  \bibinfo {author} {\bibfnamefont {G.}~\bibnamefont {Sur\'owka}}, \bibinfo
  {author} {\bibfnamefont {K.}~\bibnamefont {S\"ummerer}}, \bibinfo {author}
  {\bibfnamefont {T.}~\bibnamefont {Baumann}}, \bibinfo {author} {\bibfnamefont
  {B.}~\bibnamefont {Blank}}, \bibinfo {author} {\bibfnamefont
  {S.}~\bibnamefont {Czajkowski}}, \bibinfo {author} {\bibfnamefont
  {A.}~\bibnamefont {F\"orster}}, \bibinfo {author} {\bibfnamefont
  {M.}~\bibnamefont {Gai}}, \bibinfo {author} {\bibfnamefont {H.}~\bibnamefont
  {Geissel}}, \bibinfo {author} {\bibfnamefont {E.}~\bibnamefont {Grosse}},
  \bibinfo {author} {\bibfnamefont {M.}~\bibnamefont {Hellstr\"om}}, \bibinfo
  {author} {\bibfnamefont {P.}~\bibnamefont {Koczon}}, \bibinfo {author}
  {\bibfnamefont {B.}~\bibnamefont {Kohlmeyer}}, \bibinfo {author}
  {\bibfnamefont {R.}~\bibnamefont {Kulessa}}, \bibinfo {author} {\bibfnamefont
  {F.}~\bibnamefont {Laue}}, \bibinfo {author} {\bibfnamefont {C.}~\bibnamefont
  {Marchand}}, \bibinfo {author} {\bibfnamefont {T.}~\bibnamefont
  {Motobayashi}}, \bibinfo {author} {\bibfnamefont {H.}~\bibnamefont
  {Oeschler}}, \bibinfo {author} {\bibfnamefont {A.}~\bibnamefont {Ozawa}},
  \bibinfo {author} {\bibfnamefont {M.~S.}\ \bibnamefont {Pravikoff}}, \bibinfo
  {author} {\bibfnamefont {E.}~\bibnamefont {Schwab}}, \bibinfo {author}
  {\bibfnamefont {W.}~\bibnamefont {Schwab}}, \bibinfo {author} {\bibfnamefont
  {P.}~\bibnamefont {Senger}}, \bibinfo {author} {\bibfnamefont
  {J.}~\bibnamefont {Speer}}, \bibinfo {author} {\bibfnamefont
  {C.}~\bibnamefont {Sturm}}, \bibinfo {author} {\bibfnamefont
  {A.}~\bibnamefont {Surowiec}}, \bibinfo {author} {\bibfnamefont
  {T.}~\bibnamefont {Teranishi}}, \bibinfo {author} {\bibfnamefont
  {F.}~\bibnamefont {Uhlig}}, \bibinfo {author} {\bibfnamefont
  {A.}~\bibnamefont {Wagner}}, \bibinfo {author} {\bibfnamefont
  {W.}~\bibnamefont {Walus}},\ and\ \bibinfo {author} {\bibfnamefont {C.~A.}\
  \bibnamefont {Bertulani}},\ }\bibfield  {title} {\bibinfo {title}
  {Measurement of the coulomb dissociation of ${}^{8}\mathrm{B}$ at 254 mev
  $/$nucleon and the ${}^{8}\mathrm{B}$ solar neutrino flux},\ }\href
  {https://doi.org/10.1103/PhysRevLett.83.2910} {\bibfield  {journal} {\bibinfo
   {journal} {Phys. Rev. Lett.}\ }\textbf {\bibinfo {volume} {83}},\ \bibinfo
  {pages} {2910} (\bibinfo {year} {1999})}\BibitemShut {NoStop}%
\bibitem [{\citenamefont {Davids}\ \emph {et~al.}(2001)\citenamefont {Davids},
  \citenamefont {Anthony}, \citenamefont {Aumann}, \citenamefont {Austin},
  \citenamefont {Baumann}, \citenamefont {Bazin}, \citenamefont {Clement},
  \citenamefont {Davids}, \citenamefont {Esbensen}, \citenamefont {Lofy},
  \citenamefont {Nakamura}, \citenamefont {Sherrill},\ and\ \citenamefont
  {Yurkon}}]{Davids2001}%
  \BibitemOpen
  \bibfield  {author} {\bibinfo {author} {\bibfnamefont {B.}~\bibnamefont
  {Davids}}, \bibinfo {author} {\bibfnamefont {D.~W.}\ \bibnamefont {Anthony}},
  \bibinfo {author} {\bibfnamefont {T.}~\bibnamefont {Aumann}}, \bibinfo
  {author} {\bibfnamefont {S.~M.}\ \bibnamefont {Austin}}, \bibinfo {author}
  {\bibfnamefont {T.}~\bibnamefont {Baumann}}, \bibinfo {author} {\bibfnamefont
  {D.}~\bibnamefont {Bazin}}, \bibinfo {author} {\bibfnamefont {R.~R.~C.}\
  \bibnamefont {Clement}}, \bibinfo {author} {\bibfnamefont {C.~N.}\
  \bibnamefont {Davids}}, \bibinfo {author} {\bibfnamefont {H.}~\bibnamefont
  {Esbensen}}, \bibinfo {author} {\bibfnamefont {P.~A.}\ \bibnamefont {Lofy}},
  \bibinfo {author} {\bibfnamefont {T.}~\bibnamefont {Nakamura}}, \bibinfo
  {author} {\bibfnamefont {B.~M.}\ \bibnamefont {Sherrill}},\ and\ \bibinfo
  {author} {\bibfnamefont {J.}~\bibnamefont {Yurkon}},\ }\bibfield  {title}
  {\bibinfo {title} {${\mathit{s}}_{17}(0)$ determined from the coulomb breakup
  of 83 mev $/$nucleon $^{8}b$},\ }\href
  {https://doi.org/10.1103/PhysRevLett.86.2750} {\bibfield  {journal} {\bibinfo
   {journal} {Phys. Rev. Lett.}\ }\textbf {\bibinfo {volume} {86}},\ \bibinfo
  {pages} {2750} (\bibinfo {year} {2001})}\BibitemShut {NoStop}%
\bibitem [{\citenamefont {Baby}\ \emph {et~al.}(2003)\citenamefont {Baby},
  \citenamefont {Bordeanu}, \citenamefont {Goldring}, \citenamefont {Hass},
  \citenamefont {Weissman}, \citenamefont {Fedoseyev}, \citenamefont
  {K\"oster}, \citenamefont {Nir-El}, \citenamefont {Haquin}, \citenamefont
  {G\"aggeler},\ and\ \citenamefont {Weinreich}}]{Baby2003}%
  \BibitemOpen
  \bibfield  {author} {\bibinfo {author} {\bibfnamefont {L.~T.}\ \bibnamefont
  {Baby}}, \bibinfo {author} {\bibfnamefont {C.}~\bibnamefont {Bordeanu}},
  \bibinfo {author} {\bibfnamefont {G.}~\bibnamefont {Goldring}}, \bibinfo
  {author} {\bibfnamefont {M.}~\bibnamefont {Hass}}, \bibinfo {author}
  {\bibfnamefont {L.}~\bibnamefont {Weissman}}, \bibinfo {author}
  {\bibfnamefont {V.~N.}\ \bibnamefont {Fedoseyev}}, \bibinfo {author}
  {\bibfnamefont {U.}~\bibnamefont {K\"oster}}, \bibinfo {author}
  {\bibfnamefont {Y.}~\bibnamefont {Nir-El}}, \bibinfo {author} {\bibfnamefont
  {G.}~\bibnamefont {Haquin}}, \bibinfo {author} {\bibfnamefont {H.~W.}\
  \bibnamefont {G\"aggeler}},\ and\ \bibinfo {author} {\bibfnamefont
  {R.}~\bibnamefont {Weinreich}} (\bibinfo {collaboration} {ISOLDE
  Collaboration}),\ }\bibfield  {title} {\bibinfo {title} {Precision
  measurement of the
  $^{7}\mathrm{B}\mathrm{e}(p,\ensuremath{\gamma})^{8}\mathrm{B}$ cross section
  with an implanted $^{7}\mathrm{B}\mathrm{e}$ target},\ }\href
  {https://doi.org/10.1103/PhysRevLett.90.022501} {\bibfield  {journal}
  {\bibinfo  {journal} {Phys. Rev. Lett.}\ }\textbf {\bibinfo {volume} {90}},\
  \bibinfo {pages} {022501} (\bibinfo {year} {2003})}\BibitemShut {NoStop}%
\bibitem [{\citenamefont {Sch\"umann}\ \emph {et~al.}(2003)\citenamefont
  {Sch\"umann}, \citenamefont {Hammache}, \citenamefont {Typel}, \citenamefont
  {Uhlig}, \citenamefont {S\"ummerer}, \citenamefont {B\"ottcher},
  \citenamefont {Cortina}, \citenamefont {F\"orster}, \citenamefont {Gai},
  \citenamefont {Geissel}, \citenamefont {Greife}, \citenamefont {Iwasa},
  \citenamefont {Koczo\ifmmode~\acute{n}\else \'{n}\fi{}}, \citenamefont
  {Kohlmeyer}, \citenamefont {Kulessa}, \citenamefont {Kumagai}, \citenamefont
  {Kurz}, \citenamefont {Menzel}, \citenamefont {Motobayashi}, \citenamefont
  {Oeschler}, \citenamefont {Ozawa}, \citenamefont
  {P\l{}osko\ifmmode~\acute{n}\else \'{n}\fi{}}, \citenamefont {Prokopowicz},
  \citenamefont {Schwab}, \citenamefont {Senger}, \citenamefont {Strieder},
  \citenamefont {Sturm}, \citenamefont {Sun}, \citenamefont {Sur\'owka},
  \citenamefont {Wagner},\ and\ \citenamefont {Walu\ifmmode~\acute{s}\else
  \'{s}\fi{}}}]{Schumann2003}%
  \BibitemOpen
  \bibfield  {author} {\bibinfo {author} {\bibfnamefont {F.}~\bibnamefont
  {Sch\"umann}}, \bibinfo {author} {\bibfnamefont {F.}~\bibnamefont
  {Hammache}}, \bibinfo {author} {\bibfnamefont {S.}~\bibnamefont {Typel}},
  \bibinfo {author} {\bibfnamefont {F.}~\bibnamefont {Uhlig}}, \bibinfo
  {author} {\bibfnamefont {K.}~\bibnamefont {S\"ummerer}}, \bibinfo {author}
  {\bibfnamefont {I.}~\bibnamefont {B\"ottcher}}, \bibinfo {author}
  {\bibfnamefont {D.}~\bibnamefont {Cortina}}, \bibinfo {author} {\bibfnamefont
  {A.}~\bibnamefont {F\"orster}}, \bibinfo {author} {\bibfnamefont
  {M.}~\bibnamefont {Gai}}, \bibinfo {author} {\bibfnamefont {H.}~\bibnamefont
  {Geissel}}, \bibinfo {author} {\bibfnamefont {U.}~\bibnamefont {Greife}},
  \bibinfo {author} {\bibfnamefont {N.}~\bibnamefont {Iwasa}}, \bibinfo
  {author} {\bibfnamefont {P.}~\bibnamefont {Koczo\ifmmode~\acute{n}\else
  \'{n}\fi{}}}, \bibinfo {author} {\bibfnamefont {B.}~\bibnamefont
  {Kohlmeyer}}, \bibinfo {author} {\bibfnamefont {R.}~\bibnamefont {Kulessa}},
  \bibinfo {author} {\bibfnamefont {H.}~\bibnamefont {Kumagai}}, \bibinfo
  {author} {\bibfnamefont {N.}~\bibnamefont {Kurz}}, \bibinfo {author}
  {\bibfnamefont {M.}~\bibnamefont {Menzel}}, \bibinfo {author} {\bibfnamefont
  {T.}~\bibnamefont {Motobayashi}}, \bibinfo {author} {\bibfnamefont
  {H.}~\bibnamefont {Oeschler}}, \bibinfo {author} {\bibfnamefont
  {A.}~\bibnamefont {Ozawa}}, \bibinfo {author} {\bibfnamefont
  {M.}~\bibnamefont {P\l{}osko\ifmmode~\acute{n}\else \'{n}\fi{}}}, \bibinfo
  {author} {\bibfnamefont {W.}~\bibnamefont {Prokopowicz}}, \bibinfo {author}
  {\bibfnamefont {E.}~\bibnamefont {Schwab}}, \bibinfo {author} {\bibfnamefont
  {P.}~\bibnamefont {Senger}}, \bibinfo {author} {\bibfnamefont
  {F.}~\bibnamefont {Strieder}}, \bibinfo {author} {\bibfnamefont
  {C.}~\bibnamefont {Sturm}}, \bibinfo {author} {\bibfnamefont {Z.-Y.}\
  \bibnamefont {Sun}}, \bibinfo {author} {\bibfnamefont {G.}~\bibnamefont
  {Sur\'owka}}, \bibinfo {author} {\bibfnamefont {A.}~\bibnamefont {Wagner}},\
  and\ \bibinfo {author} {\bibfnamefont {W.}~\bibnamefont
  {Walu\ifmmode~\acute{s}\else \'{s}\fi{}}},\ }\bibfield  {title} {\bibinfo
  {title} {Coulomb dissociation of $^{8}\mathrm{B}$ and the low-energy cross
  section of the
  $^{7}\mathrm{B}\mathrm{e}(p,\ensuremath{\gamma})^{8}\mathrm{B}$ solar fusion
  reaction},\ }\href {https://doi.org/10.1103/PhysRevLett.90.232501} {\bibfield
   {journal} {\bibinfo  {journal} {Phys. Rev. Lett.}\ }\textbf {\bibinfo
  {volume} {90}},\ \bibinfo {pages} {232501} (\bibinfo {year}
  {2003})}\BibitemShut {NoStop}%
\bibitem [{\citenamefont {Junghans}\ \emph
  {et~al.}(2003{\natexlab{a}})\citenamefont {Junghans}, \citenamefont
  {Mohrmann}, \citenamefont {Snover}, \citenamefont {Steiger}, \citenamefont
  {Adelberger}, \citenamefont {Casandjian}, \citenamefont {Swanson},
  \citenamefont {Buchmann}, \citenamefont {Park}, \citenamefont {Zyuzin},\ and\
  \citenamefont {Laird}}]{Junghans2003}%
  \BibitemOpen
  \bibfield  {author} {\bibinfo {author} {\bibfnamefont {A.~R.}\ \bibnamefont
  {Junghans}}, \bibinfo {author} {\bibfnamefont {E.~C.}\ \bibnamefont
  {Mohrmann}}, \bibinfo {author} {\bibfnamefont {K.~A.}\ \bibnamefont
  {Snover}}, \bibinfo {author} {\bibfnamefont {T.~D.}\ \bibnamefont {Steiger}},
  \bibinfo {author} {\bibfnamefont {E.~G.}\ \bibnamefont {Adelberger}},
  \bibinfo {author} {\bibfnamefont {J.~M.}\ \bibnamefont {Casandjian}},
  \bibinfo {author} {\bibfnamefont {H.~E.}\ \bibnamefont {Swanson}}, \bibinfo
  {author} {\bibfnamefont {L.}~\bibnamefont {Buchmann}}, \bibinfo {author}
  {\bibfnamefont {S.~H.}\ \bibnamefont {Park}}, \bibinfo {author}
  {\bibfnamefont {A.}~\bibnamefont {Zyuzin}},\ and\ \bibinfo {author}
  {\bibfnamefont {A.~M.}\ \bibnamefont {Laird}},\ }\bibfield  {title} {\bibinfo
  {title} {Precise measurement of the
  $^{7}\mathrm{Be}(p,\ensuremath{\gamma})^{8}\mathrm{B}$ $s$ factor},\ }\href
  {https://doi.org/10.1103/PhysRevC.68.065803} {\bibfield  {journal} {\bibinfo
  {journal} {Phys. Rev. C}\ }\textbf {\bibinfo {volume} {68}},\ \bibinfo
  {pages} {065803} (\bibinfo {year} {2003}{\natexlab{a}})}\BibitemShut
  {NoStop}%
\bibitem [{\citenamefont {Sch\"umann}\ \emph {et~al.}(2006)\citenamefont
  {Sch\"umann}, \citenamefont {Typel}, \citenamefont {Hammache}, \citenamefont
  {S\"ummerer}, \citenamefont {Uhlig}, \citenamefont {B\"ottcher},
  \citenamefont {Cortina}, \citenamefont {F\"orster}, \citenamefont {Gai},
  \citenamefont {Geissel}, \citenamefont {Greife}, \citenamefont {Grosse},
  \citenamefont {Iwasa}, \citenamefont {Koczo\ifmmode~\acute{n}\else
  \'{n}\fi{}}, \citenamefont {Kohlmeyer}, \citenamefont {Kulessa},
  \citenamefont {Kumagai}, \citenamefont {Kurz}, \citenamefont {Menzel},
  \citenamefont {Motobayashi}, \citenamefont {Oeschler}, \citenamefont {Ozawa},
  \citenamefont {P\l{}osko\ifmmode~\acute{n}\else \'{n}\fi{}}, \citenamefont
  {Prokopowicz}, \citenamefont {Schwab}, \citenamefont {Senger}, \citenamefont
  {Strieder}, \citenamefont {Sturm}, \citenamefont {Sun}, \citenamefont
  {Sur\'owka}, \citenamefont {Wagner},\ and\ \citenamefont
  {Walu\ifmmode~\acute{s}\else \'{s}\fi{}}}]{Schumann2006}%
  \BibitemOpen
  \bibfield  {author} {\bibinfo {author} {\bibfnamefont {F.}~\bibnamefont
  {Sch\"umann}}, \bibinfo {author} {\bibfnamefont {S.}~\bibnamefont {Typel}},
  \bibinfo {author} {\bibfnamefont {F.}~\bibnamefont {Hammache}}, \bibinfo
  {author} {\bibfnamefont {K.}~\bibnamefont {S\"ummerer}}, \bibinfo {author}
  {\bibfnamefont {F.}~\bibnamefont {Uhlig}}, \bibinfo {author} {\bibfnamefont
  {I.}~\bibnamefont {B\"ottcher}}, \bibinfo {author} {\bibfnamefont
  {D.}~\bibnamefont {Cortina}}, \bibinfo {author} {\bibfnamefont
  {A.}~\bibnamefont {F\"orster}}, \bibinfo {author} {\bibfnamefont
  {M.}~\bibnamefont {Gai}}, \bibinfo {author} {\bibfnamefont {H.}~\bibnamefont
  {Geissel}}, \bibinfo {author} {\bibfnamefont {U.}~\bibnamefont {Greife}},
  \bibinfo {author} {\bibfnamefont {E.}~\bibnamefont {Grosse}}, \bibinfo
  {author} {\bibfnamefont {N.}~\bibnamefont {Iwasa}}, \bibinfo {author}
  {\bibfnamefont {P.}~\bibnamefont {Koczo\ifmmode~\acute{n}\else \'{n}\fi{}}},
  \bibinfo {author} {\bibfnamefont {B.}~\bibnamefont {Kohlmeyer}}, \bibinfo
  {author} {\bibfnamefont {R.}~\bibnamefont {Kulessa}}, \bibinfo {author}
  {\bibfnamefont {H.}~\bibnamefont {Kumagai}}, \bibinfo {author} {\bibfnamefont
  {N.}~\bibnamefont {Kurz}}, \bibinfo {author} {\bibfnamefont {M.}~\bibnamefont
  {Menzel}}, \bibinfo {author} {\bibfnamefont {T.}~\bibnamefont {Motobayashi}},
  \bibinfo {author} {\bibfnamefont {H.}~\bibnamefont {Oeschler}}, \bibinfo
  {author} {\bibfnamefont {A.}~\bibnamefont {Ozawa}}, \bibinfo {author}
  {\bibfnamefont {M.}~\bibnamefont {P\l{}osko\ifmmode~\acute{n}\else
  \'{n}\fi{}}}, \bibinfo {author} {\bibfnamefont {W.}~\bibnamefont
  {Prokopowicz}}, \bibinfo {author} {\bibfnamefont {E.}~\bibnamefont {Schwab}},
  \bibinfo {author} {\bibfnamefont {P.}~\bibnamefont {Senger}}, \bibinfo
  {author} {\bibfnamefont {F.}~\bibnamefont {Strieder}}, \bibinfo {author}
  {\bibfnamefont {C.}~\bibnamefont {Sturm}}, \bibinfo {author} {\bibfnamefont
  {Z.-Y.}\ \bibnamefont {Sun}}, \bibinfo {author} {\bibfnamefont
  {G.}~\bibnamefont {Sur\'owka}}, \bibinfo {author} {\bibfnamefont
  {A.}~\bibnamefont {Wagner}},\ and\ \bibinfo {author} {\bibfnamefont
  {W.}~\bibnamefont {Walu\ifmmode~\acute{s}\else \'{s}\fi{}}},\ }\bibfield
  {title} {\bibinfo {title} {Low-energy cross section of the
  $^{7}\mathrm{Be}$($p,\ensuremath{\gamma}$)$^{8}\mathrm{B}$ solar fusion
  reaction from the coulomb dissociation of $^{8}\mathrm{B}$},\ }\href
  {https://doi.org/10.1103/PhysRevC.73.015806} {\bibfield  {journal} {\bibinfo
  {journal} {Phys. Rev. C}\ }\textbf {\bibinfo {volume} {73}},\ \bibinfo
  {pages} {015806} (\bibinfo {year} {2006})}\BibitemShut {NoStop}%
\bibitem [{\citenamefont {Descouvemont}(2004)}]{Descouvemont2004}%
  \BibitemOpen
  \bibfield  {author} {\bibinfo {author} {\bibfnamefont {P.}~\bibnamefont
  {Descouvemont}},\ }\bibfield  {title} {\bibinfo {title} {Reanalysis of the
  $^{7}\mathrm{Be}(p,\ensuremath{\gamma})^{8}\mathrm{B}$ $s$ factor in a
  microscopic model},\ }\href {https://doi.org/10.1103/PhysRevC.70.065802}
  {\bibfield  {journal} {\bibinfo  {journal} {Phys. Rev. C}\ }\textbf {\bibinfo
  {volume} {70}},\ \bibinfo {pages} {065802} (\bibinfo {year}
  {2004})}\BibitemShut {NoStop}%
\bibitem [{\citenamefont {Zhang}\ \emph {et~al.}(2015)\citenamefont {Zhang},
  \citenamefont {Nollett},\ and\ \citenamefont {Phillips}}]{Zhang2015}%
  \BibitemOpen
  \bibfield  {author} {\bibinfo {author} {\bibfnamefont {X.}~\bibnamefont
  {Zhang}}, \bibinfo {author} {\bibfnamefont {K.~M.}\ \bibnamefont {Nollett}},\
  and\ \bibinfo {author} {\bibfnamefont {D.}~\bibnamefont {Phillips}},\
  }\bibfield  {title} {\bibinfo {title} {Halo effective field theory constrains
  the solar 7be + p → 8b + $\gamma$ rate},\ }\href
  {https://doi.org/https://doi.org/10.1016/j.physletb.2015.11.005} {\bibfield
  {journal} {\bibinfo  {journal} {Physics Letters B}\ }\textbf {\bibinfo
  {volume} {751}},\ \bibinfo {pages} {535} (\bibinfo {year}
  {2015})}\BibitemShut {NoStop}%
\bibitem [{\citenamefont {Machleidt}\ and\ \citenamefont
  {Entem}(2011)}]{Machleidt2011}%
  \BibitemOpen
  \bibfield  {author} {\bibinfo {author} {\bibfnamefont {R.}~\bibnamefont
  {Machleidt}}\ and\ \bibinfo {author} {\bibfnamefont {D.~R.}\ \bibnamefont
  {Entem}},\ }\bibfield  {title} {\bibinfo {title} {{Chiral effective field
  theory and nuclear forces}},\ }\href
  {https://doi.org/10.1016/j.physrep.2011.02.001} {\bibfield  {journal}
  {\bibinfo  {journal} {Phys. Rep.}\ }\textbf {\bibinfo {volume} {503}},\
  \bibinfo {pages} {1} (\bibinfo {year} {2011})}\BibitemShut {NoStop}%
\bibitem [{\citenamefont {Epelbaum}\ \emph {et~al.}(2015)\citenamefont
  {Epelbaum}, \citenamefont {Krebs},\ and\ \citenamefont
  {Mei{\ss}ner}}]{Epelbaum2015}%
  \BibitemOpen
  \bibfield  {author} {\bibinfo {author} {\bibfnamefont {E.}~\bibnamefont
  {Epelbaum}}, \bibinfo {author} {\bibfnamefont {H.}~\bibnamefont {Krebs}},\
  and\ \bibinfo {author} {\bibfnamefont {U.~G.}\ \bibnamefont {Mei{\ss}ner}},\
  }\bibfield  {title} {\bibinfo {title} {Improved chiral nucleon-nucleon
  potential up to next-to-next-to-next-to-leading order},\ }\href
  {https://doi.org/10.1140/epja/i2015-15053-8} {\bibfield  {journal} {\bibinfo
  {journal} {The European Physical Journal A}\ }\textbf {\bibinfo {volume}
  {51}},\ \bibinfo {pages} {53} (\bibinfo {year} {2015})}\BibitemShut {NoStop}%
\bibitem [{\citenamefont {Baroni}\ \emph
  {et~al.}(2013{\natexlab{a}})\citenamefont {Baroni}, \citenamefont
  {Navr\'{a}til},\ and\ \citenamefont {Quaglioni}}]{Baroni2013}%
  \BibitemOpen
  \bibfield  {author} {\bibinfo {author} {\bibfnamefont {S.}~\bibnamefont
  {Baroni}}, \bibinfo {author} {\bibfnamefont {P.}~\bibnamefont
  {Navr\'{a}til}},\ and\ \bibinfo {author} {\bibfnamefont {S.}~\bibnamefont
  {Quaglioni}},\ }\bibfield  {title} {\bibinfo {title} {{Ab Initio Description
  of the Exotic Unbound \^{}\{7\}He Nucleus}},\ }\href
  {https://doi.org/10.1103/PhysRevLett.110.022505} {\bibfield  {journal}
  {\bibinfo  {journal} {Phys. Rev. Lett.}\ }\textbf {\bibinfo {volume} {110}},\
  \bibinfo {pages} {022505} (\bibinfo {year} {2013}{\natexlab{a}})}\BibitemShut
  {NoStop}%
\bibitem [{\citenamefont {Baroni}\ \emph
  {et~al.}(2013{\natexlab{b}})\citenamefont {Baroni}, \citenamefont
  {Navr\'{a}til},\ and\ \citenamefont {Quaglioni}}]{Baroni2013a}%
  \BibitemOpen
  \bibfield  {author} {\bibinfo {author} {\bibfnamefont {S.}~\bibnamefont
  {Baroni}}, \bibinfo {author} {\bibfnamefont {P.}~\bibnamefont
  {Navr\'{a}til}},\ and\ \bibinfo {author} {\bibfnamefont {S.}~\bibnamefont
  {Quaglioni}},\ }\bibfield  {title} {\bibinfo {title} {{Unified ab initio
  approach to bound and unbound states: No-core shell model with continuum and
  its application to \^{}\{7\}He}},\ }\href
  {https://doi.org/10.1103/PhysRevC.87.034326} {\bibfield  {journal} {\bibinfo
  {journal} {Phys. Rev. C}\ }\textbf {\bibinfo {volume} {87}},\ \bibinfo
  {pages} {034326} (\bibinfo {year} {2013}{\natexlab{b}})}\BibitemShut
  {NoStop}%
\bibitem [{\citenamefont {Navr{\'{a}}til}\ \emph {et~al.}(2016)\citenamefont
  {Navr{\'{a}}til}, \citenamefont {Quaglioni}, \citenamefont {Hupin},
  \citenamefont {Romero-Redondo},\ and\ \citenamefont {Calci}}]{Navratil2016}%
  \BibitemOpen
  \bibfield  {author} {\bibinfo {author} {\bibfnamefont {P.}~\bibnamefont
  {Navr{\'{a}}til}}, \bibinfo {author} {\bibfnamefont {S.}~\bibnamefont
  {Quaglioni}}, \bibinfo {author} {\bibfnamefont {G.}~\bibnamefont {Hupin}},
  \bibinfo {author} {\bibfnamefont {C.}~\bibnamefont {Romero-Redondo}},\ and\
  \bibinfo {author} {\bibfnamefont {A.}~\bibnamefont {Calci}},\ }\bibfield
  {title} {\bibinfo {title} {Unified ab initio approaches to nuclear structure
  and reactions},\ }\href {https://doi.org/10.1088/0031-8949/91/5/053002}
  {\bibfield  {journal} {\bibinfo  {journal} {Physica Scripta}\ }\textbf
  {\bibinfo {volume} {91}},\ \bibinfo {pages} {053002} (\bibinfo {year}
  {2016})}\BibitemShut {NoStop}%
\bibitem [{\citenamefont {Barrett}\ \emph {et~al.}(2013)\citenamefont
  {Barrett}, \citenamefont {Navr\'{a}til},\ and\ \citenamefont
  {Vary}}]{Barrett2013}%
  \BibitemOpen
  \bibfield  {author} {\bibinfo {author} {\bibfnamefont {B.~R.}\ \bibnamefont
  {Barrett}}, \bibinfo {author} {\bibfnamefont {P.}~\bibnamefont
  {Navr\'{a}til}},\ and\ \bibinfo {author} {\bibfnamefont {J.~P.}\ \bibnamefont
  {Vary}},\ }\bibfield  {title} {\bibinfo {title} {{Ab initio no core shell
  model}},\ }\href {https://doi.org/10.1016/j.ppnp.2012.10.003} {\bibfield
  {journal} {\bibinfo  {journal} {Prog. Part. Nucl. Phys.}\ }\textbf {\bibinfo
  {volume} {69}},\ \bibinfo {pages} {131} (\bibinfo {year} {2013})}\BibitemShut
  {NoStop}%
\bibitem [{\citenamefont {Descouvemont}\ and\ \citenamefont
  {Baye}(2010)}]{Descouvemont2010}%
  \BibitemOpen
  \bibfield  {author} {\bibinfo {author} {\bibfnamefont {P.}~\bibnamefont
  {Descouvemont}}\ and\ \bibinfo {author} {\bibfnamefont {D.}~\bibnamefont
  {Baye}},\ }\bibfield  {title} {\bibinfo {title} {{The R -matrix theory}},\
  }\href {https://doi.org/10.1088/0034-4885/73/3/036301} {\bibfield  {journal}
  {\bibinfo  {journal} {Reports Prog. Phys.}\ }\textbf {\bibinfo {volume}
  {73}},\ \bibinfo {pages} {036301} (\bibinfo {year} {2010})}\BibitemShut
  {NoStop}%
\bibitem [{\citenamefont {Furnstahl}\ \emph {et~al.}(2015)\citenamefont
  {Furnstahl}, \citenamefont {Phillips},\ and\ \citenamefont
  {Wesolowski}}]{Furnstahl2015}%
  \BibitemOpen
  \bibfield  {author} {\bibinfo {author} {\bibfnamefont {R.~J.}\ \bibnamefont
  {Furnstahl}}, \bibinfo {author} {\bibfnamefont {D.~R.}\ \bibnamefont
  {Phillips}},\ and\ \bibinfo {author} {\bibfnamefont {S.}~\bibnamefont
  {Wesolowski}},\ }\bibfield  {title} {\bibinfo {title} {A recipe for {EFT}
  uncertainty quantification in nuclear physics},\ }\href
  {https://doi.org/10.1088/0954-3899/42/3/034028} {\bibfield  {journal}
  {\bibinfo  {journal} {Journal of Physics G: Nuclear and Particle Physics}\
  }\textbf {\bibinfo {volume} {42}},\ \bibinfo {pages} {034028} (\bibinfo
  {year} {2015})}\BibitemShut {NoStop}%
\bibitem [{\citenamefont {Melendez}\ \emph {et~al.}(2017)\citenamefont
  {Melendez}, \citenamefont {Wesolowski},\ and\ \citenamefont
  {Furnstahl}}]{Melendez2017}%
  \BibitemOpen
  \bibfield  {author} {\bibinfo {author} {\bibfnamefont {J.~A.}\ \bibnamefont
  {Melendez}}, \bibinfo {author} {\bibfnamefont {S.}~\bibnamefont
  {Wesolowski}},\ and\ \bibinfo {author} {\bibfnamefont {R.~J.}\ \bibnamefont
  {Furnstahl}},\ }\bibfield  {title} {\bibinfo {title} {Bayesian truncation
  errors in chiral effective field theory: Nucleon-nucleon observables},\
  }\href {https://doi.org/10.1103/PhysRevC.96.024003} {\bibfield  {journal}
  {\bibinfo  {journal} {Phys. Rev. C}\ }\textbf {\bibinfo {volume} {96}},\
  \bibinfo {pages} {024003} (\bibinfo {year} {2017})}\BibitemShut {NoStop}%
\bibitem [{\citenamefont {Melendez}\ \emph {et~al.}(2019)\citenamefont
  {Melendez}, \citenamefont {Furnstahl}, \citenamefont {Phillips},
  \citenamefont {Pratola},\ and\ \citenamefont {Wesolowski}}]{Melendez2019}%
  \BibitemOpen
  \bibfield  {author} {\bibinfo {author} {\bibfnamefont {J.~A.}\ \bibnamefont
  {Melendez}}, \bibinfo {author} {\bibfnamefont {R.~J.}\ \bibnamefont
  {Furnstahl}}, \bibinfo {author} {\bibfnamefont {D.~R.}\ \bibnamefont
  {Phillips}}, \bibinfo {author} {\bibfnamefont {M.~T.}\ \bibnamefont
  {Pratola}},\ and\ \bibinfo {author} {\bibfnamefont {S.}~\bibnamefont
  {Wesolowski}},\ }\bibfield  {title} {\bibinfo {title} {Quantifying correlated
  truncation errors in effective field theory},\ }\href
  {https://doi.org/10.1103/PhysRevC.100.044001} {\bibfield  {journal} {\bibinfo
   {journal} {Phys. Rev. C}\ }\textbf {\bibinfo {volume} {100}},\ \bibinfo
  {pages} {044001} (\bibinfo {year} {2019})}\BibitemShut {NoStop}%
\bibitem [{\citenamefont {Wesolowski}\ \emph {et~al.}(2021)\citenamefont
  {Wesolowski}, \citenamefont {Svensson}, \citenamefont {Ekström},
  \citenamefont {Forssén}, \citenamefont {Furnstahl}, \citenamefont
  {Melendez},\ and\ \citenamefont {Phillips}}]{Wesolowski2021}%
  \BibitemOpen
  \bibfield  {author} {\bibinfo {author} {\bibfnamefont {S.}~\bibnamefont
  {Wesolowski}}, \bibinfo {author} {\bibfnamefont {I.}~\bibnamefont
  {Svensson}}, \bibinfo {author} {\bibfnamefont {A.}~\bibnamefont {Ekström}},
  \bibinfo {author} {\bibfnamefont {C.}~\bibnamefont {Forssén}}, \bibinfo
  {author} {\bibfnamefont {R.~J.}\ \bibnamefont {Furnstahl}}, \bibinfo {author}
  {\bibfnamefont {J.~A.}\ \bibnamefont {Melendez}},\ and\ \bibinfo {author}
  {\bibfnamefont {D.~R.}\ \bibnamefont {Phillips}},\ }\bibfield  {title}
  {\bibinfo {title} {Fast \& rigorous constraints on chiral three-nucleon
  forces from few-body observables},\ }\href {https://arxiv.org/abs/2104.04441}
  {\  (\bibinfo {year} {2021})},\ \Eprint {https://arxiv.org/abs/2104.04441}
  {arXiv:2104.04441} \BibitemShut {NoStop}%
\bibitem [{\citenamefont {Ekstr\"om}\ and\ \citenamefont
  {Hagen}(2019)}]{Ekstrom2019}%
  \BibitemOpen
  \bibfield  {author} {\bibinfo {author} {\bibfnamefont {A.}~\bibnamefont
  {Ekstr\"om}}\ and\ \bibinfo {author} {\bibfnamefont {G.}~\bibnamefont
  {Hagen}},\ }\bibfield  {title} {\bibinfo {title} {Global sensitivity analysis
  of bulk properties of an atomic nucleus},\ }\href
  {https://doi.org/10.1103/PhysRevLett.123.252501} {\bibfield  {journal}
  {\bibinfo  {journal} {Phys. Rev. Lett.}\ }\textbf {\bibinfo {volume} {123}},\
  \bibinfo {pages} {252501} (\bibinfo {year} {2019})}\BibitemShut {NoStop}%
\bibitem [{\citenamefont {Kravvaris}\ \emph {et~al.}(2020)\citenamefont
  {Kravvaris}, \citenamefont {Quinlan}, \citenamefont {Quaglioni},
  \citenamefont {Wendt},\ and\ \citenamefont {Navr\'atil}}]{Kravvaris2020}%
  \BibitemOpen
  \bibfield  {author} {\bibinfo {author} {\bibfnamefont {K.}~\bibnamefont
  {Kravvaris}}, \bibinfo {author} {\bibfnamefont {K.~R.}\ \bibnamefont
  {Quinlan}}, \bibinfo {author} {\bibfnamefont {S.}~\bibnamefont {Quaglioni}},
  \bibinfo {author} {\bibfnamefont {K.~A.}\ \bibnamefont {Wendt}},\ and\
  \bibinfo {author} {\bibfnamefont {P.}~\bibnamefont {Navr\'atil}},\ }\bibfield
   {title} {\bibinfo {title} {Quantifying uncertainties in
  neutron-$\ensuremath{\alpha}$ scattering with chiral nucleon-nucleon and
  three-nucleon forces},\ }\href {https://doi.org/10.1103/PhysRevC.102.024616}
  {\bibfield  {journal} {\bibinfo  {journal} {Phys. Rev. C}\ }\textbf {\bibinfo
  {volume} {102}},\ \bibinfo {pages} {024616} (\bibinfo {year}
  {2020})}\BibitemShut {NoStop}%
\bibitem [{\citenamefont {Entem}\ and\ \citenamefont
  {Machleidt}(2003)}]{Entem2003}%
  \BibitemOpen
  \bibfield  {author} {\bibinfo {author} {\bibfnamefont {D.~R.}\ \bibnamefont
  {Entem}}\ and\ \bibinfo {author} {\bibfnamefont {R.}~\bibnamefont
  {Machleidt}},\ }\bibfield  {title} {\bibinfo {title} {{Accurate
  charge-dependent nucleon-nucleon potential at fourth order of chiral
  perturbation theory}},\ }\href {https://doi.org/10.1103/PhysRevC.68.041001}
  {\bibfield  {journal} {\bibinfo  {journal} {Phys. Rev. C}\ }\textbf {\bibinfo
  {volume} {68}},\ \bibinfo {pages} {041001(R)} (\bibinfo {year}
  {2003})}\BibitemShut {NoStop}%
\bibitem [{\citenamefont {Entem}\ \emph {et~al.}(2017)\citenamefont {Entem},
  \citenamefont {Machleidt},\ and\ \citenamefont {Nosyk}}]{Entem2017}%
  \BibitemOpen
  \bibfield  {author} {\bibinfo {author} {\bibfnamefont {D.~R.}\ \bibnamefont
  {Entem}}, \bibinfo {author} {\bibfnamefont {R.}~\bibnamefont {Machleidt}},\
  and\ \bibinfo {author} {\bibfnamefont {Y.}~\bibnamefont {Nosyk}},\ }\bibfield
   {title} {\bibinfo {title} {High-quality two-nucleon potentials up to fifth
  order of the chiral expansion},\ }\href
  {https://doi.org/10.1103/PhysRevC.96.024004} {\bibfield  {journal} {\bibinfo
  {journal} {Phys. Rev. C}\ }\textbf {\bibinfo {volume} {96}},\ \bibinfo
  {pages} {024004} (\bibinfo {year} {2017})}\BibitemShut {NoStop}%
\bibitem [{\citenamefont {van Kolck}(1994)}]{VanKolck94}%
  \BibitemOpen
  \bibfield  {author} {\bibinfo {author} {\bibfnamefont {U.}~\bibnamefont {van
  Kolck}},\ }\bibfield  {title} {\bibinfo {title} {Few-nucleon forces from
  chiral lagrangians},\ }\href {https://doi.org/10.1103/PhysRevC.49.2932}
  {\bibfield  {journal} {\bibinfo  {journal} {Phys. Rev. C}\ }\textbf {\bibinfo
  {volume} {49}},\ \bibinfo {pages} {2932} (\bibinfo {year}
  {1994})}\BibitemShut {NoStop}%
\bibitem [{\citenamefont {Navr\'{a}til}(2007)}]{Navratil2007}%
  \BibitemOpen
  \bibfield  {author} {\bibinfo {author} {\bibfnamefont {P.}~\bibnamefont
  {Navr\'{a}til}},\ }\bibfield  {title} {\bibinfo {title} {{Local three-nucleon
  interaction from chiral effective field theory}},\ }\href
  {https://doi.org/10.1007/s00601-007-0193-3} {\bibfield  {journal} {\bibinfo
  {journal} {Few-Body Syst.}\ }\textbf {\bibinfo {volume} {41}},\ \bibinfo
  {pages} {117} (\bibinfo {year} {2007})}\BibitemShut {NoStop}%
\bibitem [{\citenamefont {Gazit}\ \emph {et~al.}(2019)\citenamefont {Gazit},
  \citenamefont {Quaglioni},\ and\ \citenamefont {Navr\'atil}}]{Gazit2019}%
  \BibitemOpen
  \bibfield  {author} {\bibinfo {author} {\bibfnamefont {D.}~\bibnamefont
  {Gazit}}, \bibinfo {author} {\bibfnamefont {S.}~\bibnamefont {Quaglioni}},\
  and\ \bibinfo {author} {\bibfnamefont {P.}~\bibnamefont {Navr\'atil}},\
  }\bibfield  {title} {\bibinfo {title} {Erratum: Three-nucleon low-energy
  constants from the consistency of interactions and currents in chiral
  effective field theory [phys. rev. lett. 103, 102502 (2009)]},\ }\href
  {https://doi.org/10.1103/PhysRevLett.122.029901} {\bibfield  {journal}
  {\bibinfo  {journal} {Phys. Rev. Lett.}\ }\textbf {\bibinfo {volume} {122}},\
  \bibinfo {pages} {029901(E)} (\bibinfo {year} {2019})}\BibitemShut {NoStop}%
\bibitem [{\citenamefont {Som\`a}\ \emph {et~al.}(2020)\citenamefont {Som\`a},
  \citenamefont {Navr\'atil}, \citenamefont {Raimondi}, \citenamefont
  {Barbieri},\ and\ \citenamefont {Duguet}}]{Soma2020}%
  \BibitemOpen
  \bibfield  {author} {\bibinfo {author} {\bibfnamefont {V.}~\bibnamefont
  {Som\`a}}, \bibinfo {author} {\bibfnamefont {P.}~\bibnamefont {Navr\'atil}},
  \bibinfo {author} {\bibfnamefont {F.}~\bibnamefont {Raimondi}}, \bibinfo
  {author} {\bibfnamefont {C.}~\bibnamefont {Barbieri}},\ and\ \bibinfo
  {author} {\bibfnamefont {T.}~\bibnamefont {Duguet}},\ }\bibfield  {title}
  {\bibinfo {title} {Novel chiral hamiltonian and observables in light and
  medium-mass nuclei},\ }\href {https://doi.org/10.1103/PhysRevC.101.014318}
  {\bibfield  {journal} {\bibinfo  {journal} {Phys. Rev. C}\ }\textbf {\bibinfo
  {volume} {101}},\ \bibinfo {pages} {014318} (\bibinfo {year}
  {2020})}\BibitemShut {NoStop}%
\bibitem [{\citenamefont {Girlanda}\ \emph {et~al.}(2011)\citenamefont
  {Girlanda}, \citenamefont {Kievsky},\ and\ \citenamefont
  {Viviani}}]{Girlanda2011}%
  \BibitemOpen
  \bibfield  {author} {\bibinfo {author} {\bibfnamefont {L.}~\bibnamefont
  {Girlanda}}, \bibinfo {author} {\bibfnamefont {A.}~\bibnamefont {Kievsky}},\
  and\ \bibinfo {author} {\bibfnamefont {M.}~\bibnamefont {Viviani}},\
  }\bibfield  {title} {\bibinfo {title} {Subleading contributions to the
  three-nucleon contact interaction},\ }\href
  {https://doi.org/10.1103/PhysRevC.84.014001} {\bibfield  {journal} {\bibinfo
  {journal} {Phys. Rev. C}\ }\textbf {\bibinfo {volume} {84}},\ \bibinfo
  {pages} {014001} (\bibinfo {year} {2011})}\BibitemShut {NoStop}%
\bibitem [{\citenamefont {Jurgenson}\ \emph {et~al.}(2009)\citenamefont
  {Jurgenson}, \citenamefont {Navr\'{a}til},\ and\ \citenamefont
  {Furnstahl}}]{Jurgenson2009}%
  \BibitemOpen
  \bibfield  {author} {\bibinfo {author} {\bibfnamefont {E.~D.}\ \bibnamefont
  {Jurgenson}}, \bibinfo {author} {\bibfnamefont {P.}~\bibnamefont
  {Navr\'{a}til}},\ and\ \bibinfo {author} {\bibfnamefont {R.~J.}\ \bibnamefont
  {Furnstahl}},\ }\bibfield  {title} {\bibinfo {title} {{Evolution of Nuclear
  Many-Body Forces with the Similarity Renormalization Group}},\ }\href
  {https://doi.org/10.1103/PhysRevLett.103.082501} {\bibfield  {journal}
  {\bibinfo  {journal} {Phys. Rev. Lett.}\ }\textbf {\bibinfo {volume} {103}},\
  \bibinfo {pages} {082501} (\bibinfo {year} {2009})}\BibitemShut {NoStop}%
\bibitem [{\citenamefont {McCracken}\ \emph {et~al.}(2021)\citenamefont
  {McCracken}, \citenamefont {Navr\'atil}, \citenamefont {McCoy}, \citenamefont
  {Quaglioni},\ and\ \citenamefont {Hupin}}]{PhysRevC.103.035801}%
  \BibitemOpen
  \bibfield  {author} {\bibinfo {author} {\bibfnamefont {C.}~\bibnamefont
  {McCracken}}, \bibinfo {author} {\bibfnamefont {P.}~\bibnamefont
  {Navr\'atil}}, \bibinfo {author} {\bibfnamefont {A.}~\bibnamefont {McCoy}},
  \bibinfo {author} {\bibfnamefont {S.}~\bibnamefont {Quaglioni}},\ and\
  \bibinfo {author} {\bibfnamefont {G.}~\bibnamefont {Hupin}},\ }\bibfield
  {title} {\bibinfo {title} {Microscopic investigation of the
  $^{8}\mathrm{Li}(n,\ensuremath{\gamma})^{9}\mathrm{Li}$ reaction},\ }\href
  {https://doi.org/10.1103/PhysRevC.103.035801} {\bibfield  {journal} {\bibinfo
   {journal} {Phys. Rev. C}\ }\textbf {\bibinfo {volume} {103}},\ \bibinfo
  {pages} {035801} (\bibinfo {year} {2021})}\BibitemShut {NoStop}%
\bibitem [{\citenamefont {Navr\'{a}til}\ \emph {et~al.}(2011)\citenamefont
  {Navr\'{a}til}, \citenamefont {Roth},\ and\ \citenamefont
  {Quaglioni}}]{Navratil2011a}%
  \BibitemOpen
  \bibfield  {author} {\bibinfo {author} {\bibfnamefont {P.}~\bibnamefont
  {Navr\'{a}til}}, \bibinfo {author} {\bibfnamefont {R.}~\bibnamefont {Roth}},\
  and\ \bibinfo {author} {\bibfnamefont {S.}~\bibnamefont {Quaglioni}},\
  }\bibfield  {title} {\bibinfo {title} {{Ab initio many-body calculation of
  the radiative capture}},\ }\href
  {https://doi.org/10.1016/j.physletb.2011.09.079} {\bibfield  {journal}
  {\bibinfo  {journal} {Phys. Lett. B}\ }\textbf {\bibinfo {volume} {704}},\
  \bibinfo {pages} {379} (\bibinfo {year} {2011})}\BibitemShut {NoStop}%
\bibitem [{\citenamefont {Navr\'{a}til}\ and\ \citenamefont
  {Quaglioni}(2011)}]{Navratil2011}%
  \BibitemOpen
  \bibfield  {author} {\bibinfo {author} {\bibfnamefont {P.}~\bibnamefont
  {Navr\'{a}til}}\ and\ \bibinfo {author} {\bibfnamefont {S.}~\bibnamefont
  {Quaglioni}},\ }\bibfield  {title} {\bibinfo {title} {{Ab initio many-body
  calculations of deuteron-\^{}\{4\}He scattering and \^{}\{6\}Li states}},\
  }\href {https://doi.org/10.1103/PhysRevC.83.044609} {\bibfield  {journal}
  {\bibinfo  {journal} {Phys. Rev. C}\ }\textbf {\bibinfo {volume} {83}},\
  \bibinfo {pages} {044609} (\bibinfo {year} {2011})}\BibitemShut {NoStop}%
\bibitem [{\citenamefont {Paneru}\ \emph {et~al.}(2019)\citenamefont {Paneru},
  \citenamefont {Brune}, \citenamefont {Giri}, \citenamefont {Livesay},
  \citenamefont {Greife}, \citenamefont {Blackmon}, \citenamefont {Bardayan},
  \citenamefont {Chipps}, \citenamefont {Davids}, \citenamefont {Connolly},
  \citenamefont {Chae}, \citenamefont {Champagne}, \citenamefont {Deibel},
  \citenamefont {Jones}, \citenamefont {Johnson}, \citenamefont {Kozub},
  \citenamefont {Ma}, \citenamefont {Nesaraja}, \citenamefont {Pain},
  \citenamefont {Sarazin}, \citenamefont {Shriner}, \citenamefont {Stracener},
  \citenamefont {Smith}, \citenamefont {Thomas}, \citenamefont {Visser},\ and\
  \citenamefont {Wrede}}]{PhysRevC.99.045807}%
  \BibitemOpen
  \bibfield  {author} {\bibinfo {author} {\bibfnamefont {S.~N.}\ \bibnamefont
  {Paneru}}, \bibinfo {author} {\bibfnamefont {C.~R.}\ \bibnamefont {Brune}},
  \bibinfo {author} {\bibfnamefont {R.}~\bibnamefont {Giri}}, \bibinfo {author}
  {\bibfnamefont {R.~J.}\ \bibnamefont {Livesay}}, \bibinfo {author}
  {\bibfnamefont {U.}~\bibnamefont {Greife}}, \bibinfo {author} {\bibfnamefont
  {J.~C.}\ \bibnamefont {Blackmon}}, \bibinfo {author} {\bibfnamefont {D.~W.}\
  \bibnamefont {Bardayan}}, \bibinfo {author} {\bibfnamefont {K.~A.}\
  \bibnamefont {Chipps}}, \bibinfo {author} {\bibfnamefont {B.}~\bibnamefont
  {Davids}}, \bibinfo {author} {\bibfnamefont {D.~S.}\ \bibnamefont
  {Connolly}}, \bibinfo {author} {\bibfnamefont {K.~Y.}\ \bibnamefont {Chae}},
  \bibinfo {author} {\bibfnamefont {A.~E.}\ \bibnamefont {Champagne}}, \bibinfo
  {author} {\bibfnamefont {C.}~\bibnamefont {Deibel}}, \bibinfo {author}
  {\bibfnamefont {K.~L.}\ \bibnamefont {Jones}}, \bibinfo {author}
  {\bibfnamefont {M.~S.}\ \bibnamefont {Johnson}}, \bibinfo {author}
  {\bibfnamefont {R.~L.}\ \bibnamefont {Kozub}}, \bibinfo {author}
  {\bibfnamefont {Z.}~\bibnamefont {Ma}}, \bibinfo {author} {\bibfnamefont
  {C.~D.}\ \bibnamefont {Nesaraja}}, \bibinfo {author} {\bibfnamefont {S.~D.}\
  \bibnamefont {Pain}}, \bibinfo {author} {\bibfnamefont {F.}~\bibnamefont
  {Sarazin}}, \bibinfo {author} {\bibfnamefont {J.~F.}\ \bibnamefont
  {Shriner}}, \bibinfo {author} {\bibfnamefont {D.~W.}\ \bibnamefont
  {Stracener}}, \bibinfo {author} {\bibfnamefont {M.~S.}\ \bibnamefont
  {Smith}}, \bibinfo {author} {\bibfnamefont {J.~S.}\ \bibnamefont {Thomas}},
  \bibinfo {author} {\bibfnamefont {D.~W.}\ \bibnamefont {Visser}},\ and\
  \bibinfo {author} {\bibfnamefont {C.}~\bibnamefont {Wrede}},\ }\bibfield
  {title} {\bibinfo {title} {$s$-wave scattering lengths for the
  $^{7}\mathrm{Be}+p$ system from an $r$-matrix analysis},\ }\href
  {https://doi.org/10.1103/PhysRevC.99.045807} {\bibfield  {journal} {\bibinfo
  {journal} {Phys. Rev. C}\ }\textbf {\bibinfo {volume} {99}},\ \bibinfo
  {pages} {045807} (\bibinfo {year} {2019})}\BibitemShut {NoStop}%
\bibitem [{\citenamefont {Robertson}(1973)}]{Robertson1973}%
  \BibitemOpen
  \bibfield  {author} {\bibinfo {author} {\bibfnamefont {R.~G.~H.}\
  \bibnamefont {Robertson}},\ }\bibfield  {title} {\bibinfo {title} {Proton
  capture by $^{7}\mathrm{Be}$ and the solar neutrino problem},\ }\href
  {https://doi.org/10.1103/PhysRevC.7.543} {\bibfield  {journal} {\bibinfo
  {journal} {Phys. Rev. C}\ }\textbf {\bibinfo {volume} {7}},\ \bibinfo {pages}
  {543} (\bibinfo {year} {1973})}\BibitemShut {NoStop}%
\bibitem [{\citenamefont {Schuster}\ \emph {et~al.}(2014)\citenamefont
  {Schuster}, \citenamefont {Quaglioni}, \citenamefont {Johnson}, \citenamefont
  {Jurgenson},\ and\ \citenamefont {Navr\'atil}}]{Schuster2014}%
  \BibitemOpen
  \bibfield  {author} {\bibinfo {author} {\bibfnamefont {M.~D.}\ \bibnamefont
  {Schuster}}, \bibinfo {author} {\bibfnamefont {S.}~\bibnamefont {Quaglioni}},
  \bibinfo {author} {\bibfnamefont {C.~W.}\ \bibnamefont {Johnson}}, \bibinfo
  {author} {\bibfnamefont {E.~D.}\ \bibnamefont {Jurgenson}},\ and\ \bibinfo
  {author} {\bibfnamefont {P.}~\bibnamefont {Navr\'atil}},\ }\bibfield  {title}
  {\bibinfo {title} {Operator evolution for ab initio theory of light nuclei},\
  }\href {https://doi.org/10.1103/PhysRevC.90.011301} {\bibfield  {journal}
  {\bibinfo  {journal} {Phys. Rev. C}\ }\textbf {\bibinfo {volume} {90}},\
  \bibinfo {pages} {011301(R)} (\bibinfo {year} {2014})}\BibitemShut {NoStop}%
\bibitem [{\citenamefont {Baye}\ and\ \citenamefont
  {Brainis}(2000)}]{PhysRevC.61.025801}%
  \BibitemOpen
  \bibfield  {author} {\bibinfo {author} {\bibfnamefont {D.}~\bibnamefont
  {Baye}}\ and\ \bibinfo {author} {\bibfnamefont {E.}~\bibnamefont {Brainis}},\
  }\bibfield  {title} {\bibinfo {title} {Zero-energy determination of the
  astrophysical s factor and effective-range expansions},\ }\href
  {https://doi.org/10.1103/PhysRevC.61.025801} {\bibfield  {journal} {\bibinfo
  {journal} {Phys. Rev. C}\ }\textbf {\bibinfo {volume} {61}},\ \bibinfo
  {pages} {025801} (\bibinfo {year} {2000})}\BibitemShut {NoStop}%
\bibitem [{\citenamefont {Audi}\ \emph {et~al.}(2012)\citenamefont {Audi},
  \citenamefont {Kondev}, \citenamefont {Wang}, \citenamefont {Pfeiffer},
  \citenamefont {Sun}, \citenamefont {Blachot},\ and\ \citenamefont
  {MacCormick}}]{Audi2012}%
  \BibitemOpen
  \bibfield  {author} {\bibinfo {author} {\bibfnamefont {G.}~\bibnamefont
  {Audi}}, \bibinfo {author} {\bibfnamefont {F.~G.}\ \bibnamefont {Kondev}},
  \bibinfo {author} {\bibfnamefont {M.}~\bibnamefont {Wang}}, \bibinfo {author}
  {\bibfnamefont {B.}~\bibnamefont {Pfeiffer}}, \bibinfo {author}
  {\bibfnamefont {X.}~\bibnamefont {Sun}}, \bibinfo {author} {\bibfnamefont
  {J.}~\bibnamefont {Blachot}},\ and\ \bibinfo {author} {\bibfnamefont
  {M.}~\bibnamefont {MacCormick}},\ }\bibfield  {title} {\bibinfo {title} {{The
  Nubase2012 evaluation of nuclear properties}},\ }\href
  {https://doi.org/10.1088/1674-1137/36/12/001} {\bibfield  {journal} {\bibinfo
   {journal} {Chinese Phys. C}\ }\textbf {\bibinfo {volume} {36}},\ \bibinfo
  {pages} {1157} (\bibinfo {year} {2012})}\BibitemShut {NoStop}%
\bibitem [{\citenamefont {Hupin}\ \emph {et~al.}(2019)\citenamefont {Hupin},
  \citenamefont {Quaglioni},\ and\ \citenamefont {Navr{\'a}til}}]{Hupin2019}%
  \BibitemOpen
  \bibfield  {author} {\bibinfo {author} {\bibfnamefont {G.}~\bibnamefont
  {Hupin}}, \bibinfo {author} {\bibfnamefont {S.}~\bibnamefont {Quaglioni}},\
  and\ \bibinfo {author} {\bibfnamefont {P.}~\bibnamefont {Navr{\'a}til}},\
  }\bibfield  {title} {\bibinfo {title} {Ab initio predictions for polarized
  deuterium-tritium thermonuclear fusion},\ }\href
  {https://doi.org/10.1038/s41467-018-08052-6} {\bibfield  {journal} {\bibinfo
  {journal} {Nature Communications}\ }\textbf {\bibinfo {volume} {10}},\
  \bibinfo {pages} {351} (\bibinfo {year} {2019})}\BibitemShut {NoStop}%
\bibitem [{\citenamefont {Junghans}\ \emph
  {et~al.}(2003{\natexlab{b}})\citenamefont {Junghans}, \citenamefont
  {Mohrmann}, \citenamefont {Snover}, \citenamefont {Steiger}, \citenamefont
  {Adelberger}, \citenamefont {Casandjian}, \citenamefont {Swanson},
  \citenamefont {Buchmann}, \citenamefont {Park}, \citenamefont {Zyuzin},\ and\
  \citenamefont {Laird}}]{PhysRevC.68.065803}%
  \BibitemOpen
  \bibfield  {author} {\bibinfo {author} {\bibfnamefont {A.~R.}\ \bibnamefont
  {Junghans}}, \bibinfo {author} {\bibfnamefont {E.~C.}\ \bibnamefont
  {Mohrmann}}, \bibinfo {author} {\bibfnamefont {K.~A.}\ \bibnamefont
  {Snover}}, \bibinfo {author} {\bibfnamefont {T.~D.}\ \bibnamefont {Steiger}},
  \bibinfo {author} {\bibfnamefont {E.~G.}\ \bibnamefont {Adelberger}},
  \bibinfo {author} {\bibfnamefont {J.~M.}\ \bibnamefont {Casandjian}},
  \bibinfo {author} {\bibfnamefont {H.~E.}\ \bibnamefont {Swanson}}, \bibinfo
  {author} {\bibfnamefont {L.}~\bibnamefont {Buchmann}}, \bibinfo {author}
  {\bibfnamefont {S.~H.}\ \bibnamefont {Park}}, \bibinfo {author}
  {\bibfnamefont {A.}~\bibnamefont {Zyuzin}},\ and\ \bibinfo {author}
  {\bibfnamefont {A.~M.}\ \bibnamefont {Laird}},\ }\bibfield  {title} {\bibinfo
  {title} {Precise measurement of the
  $^{7}\mathrm{Be}(p,\ensuremath{\gamma})^{8}\mathrm{B}$ $s$ factor},\ }\href
  {https://doi.org/10.1103/PhysRevC.68.065803} {\bibfield  {journal} {\bibinfo
  {journal} {Phys. Rev. C}\ }\textbf {\bibinfo {volume} {68}},\ \bibinfo
  {pages} {065803} (\bibinfo {year} {2003}{\natexlab{b}})}\BibitemShut
  {NoStop}%
\bibitem [{\citenamefont {Xu}\ \emph {et~al.}(1994)\citenamefont {Xu},
  \citenamefont {Gagliardi}, \citenamefont {Tribble}, \citenamefont
  {Mukhamedzhanov},\ and\ \citenamefont {Timofeyuk}}]{Xu1994}%
  \BibitemOpen
  \bibfield  {author} {\bibinfo {author} {\bibfnamefont {H.~M.}\ \bibnamefont
  {Xu}}, \bibinfo {author} {\bibfnamefont {C.~A.}\ \bibnamefont {Gagliardi}},
  \bibinfo {author} {\bibfnamefont {R.~E.}\ \bibnamefont {Tribble}}, \bibinfo
  {author} {\bibfnamefont {A.~M.}\ \bibnamefont {Mukhamedzhanov}},\ and\
  \bibinfo {author} {\bibfnamefont {N.~K.}\ \bibnamefont {Timofeyuk}},\
  }\bibfield  {title} {\bibinfo {title} {Overall normalization of the
  astrophysical $s$ factor and the nuclear vertex constant for
  $^{7}\mathrm{Be}{(p,\ensuremath{\gamma})}^{8}\mathrm{B}$ reactions},\ }\href
  {https://doi.org/10.1103/PhysRevLett.73.2027} {\bibfield  {journal} {\bibinfo
   {journal} {Phys. Rev. Lett.}\ }\textbf {\bibinfo {volume} {73}},\ \bibinfo
  {pages} {2027} (\bibinfo {year} {1994})}\BibitemShut {NoStop}%
\bibitem [{\citenamefont {Baye}(2000)}]{Baye2000b}%
  \BibitemOpen
  \bibfield  {author} {\bibinfo {author} {\bibfnamefont {D.}~\bibnamefont
  {Baye}},\ }\bibfield  {title} {\bibinfo {title} {Behavior of the
  ${}^{7}\mathrm{Be}(p,\ensuremath{\gamma}{)}^{8}\mathrm{B}$ astrophysical s
  factor near zero energy},\ }\href
  {https://doi.org/10.1103/PhysRevC.62.065803} {\bibfield  {journal} {\bibinfo
  {journal} {Phys. Rev. C}\ }\textbf {\bibinfo {volume} {62}},\ \bibinfo
  {pages} {065803} (\bibinfo {year} {2000})}\BibitemShut {NoStop}%
\end{thebibliography}
\end{document}